\documentclass[sn-mathphys]{sn-jnl}

\usepackage{mathrsfs}
\usepackage{amsmath}
\usepackage{amsfonts}
\usepackage{amsthm}
\usepackage{color}
\usepackage{xcolor}
\usepackage{graphicx}
\usepackage{tabularx}

\usepackage{bm}

\usepackage{caption}
\usepackage{subcaption}

\usepackage{float}
\usepackage{url}
\usepackage{multicol}
\usepackage{txfonts}
\usepackage{setspace}
\usepackage{arydshln}
\usepackage{enumitem}
\usepackage{lipsum}
\usepackage{siunitx,booktabs}
\usepackage{nth}
\usepackage{accents} 

\labelformat{enumi}{#1}
\labelformat{enumii}{#1}
\labelformat{enumiii}{#1}
\labelformat{enumiv}{#1}

\theoremstyle{thmstyleone}%
\theoremstyle{thmstyletwo}%

\theoremstyle{thmstylethree}%

\usepackage{listings}
\definecolor{light-gray}{gray}{0.95}
\newcommand{\authorname}[2]{\mbox{#1~#2}}

\begin{document}

\title[Comparative analysis of RIS and IB methods for aortic valve simulation]{Comparative analysis of resistive immersed surface and immersed boundary methods for aortic valve simulation}


\author[1]{\authorname{\fnm{Han}}{\sur{Zhao}}}

\author[2,3,4]{\authorname{\fnm{Alexander D.}}{\sur{Kaiser}}}

\author[5]{\authorname{\fnm{Fanwei}}{\sur{Kong}}}

\author[6]{\authorname{\fnm{Aaron L.}}{\sur{Brown}}}

\author[6]{\authorname{\fnm{Zinan}}{\sur{Hu}}}

\author[1]{\authorname{\fnm{David}}{\sur{Codoni}}}

\author[1]{\authorname{\fnm{Sujal}}{\sur{Dave}}}

\author*[1,3,6,7,8]{\authorname{\fnm{Alison L.}}{\sur{Marsden}}}
\email{amarsden@stanford.edu}

\affil[1]{\orgdiv{Department of Pediatrics (Cardiology)}, \orgname{Stanford University}, \orgaddress{\city{Stanford}, \state{CA}, \country{USA}}}

\affil[2]{\orgdiv{Department of Cardiothoracic Surgery}, \orgname{Stanford University}, \orgaddress{\city{Stanford}, \state{CA}, \country{USA}}}

\affil[3]{\orgdiv{Stanford Cardiovascular Institute}, \orgname{Stanford University}, \orgaddress{\city{Stanford}, \state{CA}, \country{USA}}}

\affil[4]{\orgdiv{Maternal \& Child Health Research Institute}, \orgname{Stanford University}, \orgaddress{\city{Stanford}, \state{CA}, \country{USA}}}

\affil[5]{\orgdiv{Department of Mechanical Engineering and Materials Science}, \orgname{Washington University}, \orgaddress{\city{St. Louis}, \state{MO}, \country{USA}}}

\affil[6]{\orgdiv{Department of Mechanical Engineering}, \orgname{Stanford University}, \orgaddress{\city{Stanford}, \state{CA}, \country{USA}}}

\affil[7]{\orgdiv{Department of Bioengineering}, \orgname{Stanford University}, \orgaddress{\city{Stanford}, \state{CA}, \country{USA}}}

\affil[8]{\orgdiv{Institute for Computational and Mathematical Engineering}, \orgname{Stanford University}, \orgaddress{\city{Stanford}, \state{CA}, \country{USA}}}

\abstract{Numerical modeling of aortic valve dynamics is essential for understanding the complex fluid--structure interaction (FSI) governing valve biomechanics in health and disease. Immersed methods provide a flexible computational framework for simulating the large deformations of valve leaflets and associated blood flow without requiring body-fitted meshes. Among these approaches, the Resistive Immersed Surface (RIS) and Immersed Boundary (IB) methods are widely used. However, systematic comparative analysis of these methods for realistic aortic valve simulations has not been performed. In this work, we compare a prescribed-kinematics RIS workflow implemented in SimVascular's svMultiPhysics solver with a fully coupled IB workflow using IBAMR for trileaflet and bicuspid aortic valve configurations. The RIS method represents the valve as a surface with prescribed kinematics embedded in the fluid domain and introduces a penalty force that drives the surrounding fluid velocity toward the prescribed leaflet velocity. This formulation reduces modeling complexity and provides useful hemodynamic predictions when representative leaflet kinematics are available. In contrast, the IB method models the leaflets as elastic structures fully immersed in the fluid domain and resolves leaflet deformation through fully coupled two-way FSI. The study focuses on the extent to which RIS reproduces bulk hemodynamic features and transvalvular pressure gradients. Results show that the RIS method captures the large-scale flow structures and predicts the mean transvalvular pressure gradient with a relative error within 15\% of the fully coupled IB simulation, improving to within 5\% when inlet boundary conditions are matched, while reducing computational cost by approximately 60\%.}




\maketitle



\section{Introduction} \label{sec:intro}

Aortic valve disease, including aortic stenosis and aortic regurgitation arising from calcific degeneration, rheumatic disease, or congenital malformations, represents a significant health burden in the United States. Valvular heart disease is estimated to affect approximately 2.5\% of the adult population~\cite{nkomo2006burden, iung2011epidemiology}. The aortic valve regulates blood flow from the left ventricle into the aorta, and its dysfunction can lead to reduced cardiac output, elevated ventricular pressure load, and heart failure if left untreated~\cite{carabello2009aortic,carabello2013introduction}. Clinical treatment ranges from surgical valve repair or replacement to minimally invasive transcatheter aortic valve replacement (TAVR), with outcomes that improve the hemodynamic environment created by the diseased valve~\cite{writing20212020}. Computational simulation provides a means to predict the outcomes of various treatment scenarios in a controlled manner, providing realistic blood velocity and pressure under clinically relevant conditions that can be tuned to patient data~\cite{taylor2009patient, marsden2015multiscale}. Such simulations have been used to assess transvalvular pressure gradients and regurgitant flow in diseased valves~\cite{franke2020towards}, to evaluate the hemodynamic performance of prosthetic valve designs~\cite{xu2021computational, hsu2018immersogeometric, wu2018anisotropic}, and to support surgical planning by predicting post-intervention flow patterns~\cite{kaiser2021design, kaiser2024simulation, kaiser2025simulation}. Computational modeling of aortic valve hemodynamics therefore plays an important role in both understanding valve disease and developing improved treatments.

The aortic valve undergoes large and rapid deformations under pulsatile blood flow during each cardiac cycle. Valve closure also involves strong geometric nonlinearity and large pressure gradients, making accurate simulation a challenging fluid--structure interaction (FSI) problem. Conventional body-fitted FSI approaches, such as the arbitrary Lagrangian--Eulerian (ALE) method, require the fluid mesh to deform with the moving leaflet boundaries. For heart valves, the large magnitude and rapidity of leaflet motion make it difficult to maintain mesh quality over a full cardiac cycle without remeshing, and leaflet coaptation during valve closure introduces topological changes that conventional body-fitted meshes cannot accommodate without additional treatment, thereby increasing algorithmic complexity and computational cost~\cite{johnson1994mesh, bazilevs2008isogeometric, votta2013toward, gerosa2024mechanically}. Immersed-type methods address this challenge by embedding the valve structure within a nonconforming background fluid mesh, thereby avoiding mesh deformation altogether.

Immersed-type methods on nonconforming meshes are widely used for heart valve simulation. The classical Immersed Boundary (IB) method~\cite{peskin1972flow, peskin1977numerical, peskin2002immersed} models the valve leaflets as elastic Lagrangian structures fully immersed in an Eulerian fluid domain in a two-way coupled manner. Elastic forces from the structure are spread to the fluid grid through a regularized Dirac delta kernel, while the structural velocity is obtained by interpolating the fluid velocity to the Lagrangian points. This two-way coupling allows leaflet motion to respond directly to fluid loading. Over the past several decades, the IB method has undergone substantial development, including advances in adaptive mesh refinement~\cite{griffith2007adaptive, roma1999adaptive}, kernel function accuracy~\cite{bao2015gaussian, bale2021one, gruninger2025composite}, and realistic heart valve simulations~\cite{griffith2009simulating, kaiser2022controlled}, with increasing clinical relevance~\cite{kaiser2024simulation, kaiser2025simulation}. Valve models based on the IB method have been validated by direct comparisons against in vitro and ex vivo experimental data, including in vitro 4D flow MRI of bioprosthetic pulmonary valves~\cite{kaiser2023comparison}, pulse duplicator measurements of bioprosthetic valves~\cite{lee2020fluid}, and ex vivo testing of bicuspidization repairs~\cite{choi2024combined}, which provide quantitative benchmarks for transvalvular pressure gradients, flow, and leaflet kinematics.  A widely used open-source scalable implementation of the IB method is provided by the IBAMR (Immersed Boundary Method Adaptive Mesh Refinement) software~\cite{griffith2017ibamr}.  Other variants of immersed methods include immersed finite element formulations~\cite{zhang2004immersed, liu2006immersed, wang2013modified}, fictitious domain approaches~\cite{glowinski1999distributed, boffi2015finite, black2025immersed}, and immersogeometric analysis (IMGA)~\cite{hsu2014fluid, kamensky2015immersogeometric, kamensky2017immersogeometric, hsu2018immersogeometric, wu2018anisotropic}, in which NURBS-based isogeometric leaflet representations are coupled to a nonconforming background fluid mesh. An open-source implementation of IMGA is available through the CouDALFISh library~\cite{kamensky2019tigar, Kamensky2021, zhao2022open, neighbor2023leveraging}. A recent open-source framework for heart valve simulation~\cite{black2026open}, implemented based on MFEM~\cite{anderson2021mfem} and FEBio~\cite{maas2012febio}, adopts a fictitious-domain approach. An essential distinction across these immersed-type approaches lies in the type of background fluid mesh. The classical IB method is formulated on structured Cartesian grids, whereas the immersed-type variants, including the fictitious-domain method, immersed finite element method, and IMGA, support unstructured background meshes.

In contrast, the Resistive Immersed Surface (RIS) method~\cite{fernandez2008numerical, astorino2012robust, fedele2017patient, this2020augmented, zingaro2023modeling} represents the valve leaflets by surface meshes whose motion is prescribed a priori. A resistive penalty force, localized near the valve surface through a smoothed Dirac delta function, drives the fluid velocity toward the prescribed leaflet velocity. Because the leaflet kinematics are supplied as input rather than computed from fluid loading, the RIS formulation is one-way coupled and requires only the solution of the fluid equations. The prescribed kinematics can be obtained from medical imaging, structural analyses, or prior FSI simulations, which simplifies the equations of motion and modeling complexity. The RIS method can be implemented on either structured or unstructured grids, and the finite element formulation used in the present work is computed on unstructured meshes. Recent applications of RIS include patient-specific aortic valve simulations~\cite{fedele2017patient}, multiphysics simulations of the human left heart~\cite{bucelli2023mathematical, bucelli2025coupling, ruz2025left}, and implementation in the open-source framework $\text{life}^\text{x}$-cfd~\cite{africa2024lifex}.

Although both the IB and RIS methods have been applied independently to aortic valve simulation, systematic side-by-side comparisons on the same realistic geometries and under the same physiological flow conditions remain limited. Such a comparison is practically useful because the two methods represent different trade-offs between computational cost and physical fidelity, and understanding when the simpler RIS approach is sufficient can provide concrete guidance for method selection. In this work, we present a direct comparison of the RIS and IB methods for trileaflet and bicuspid aortic valve configurations in a realistic aortic geometry. The RIS method is implemented in the open-source finite element solver svMultiPhysics~\cite{zhu2022svfsi, svmultiphysics-code}, which uses a residual-based variational multiscale (VMS) formulation~\cite{bazilevs2007variational, liu2018unified} and is part of the SimVascular platform~\cite{updegrove2017simvascular}. The IB simulations are performed using IBAMR~\cite{griffith2017ibamr, ibamr-code}, which solves the incompressible Navier--Stokes equations on a block-structured adaptive Cartesian grid. Physiologically realistic boundary conditions are imposed through lumped-parameter models~\cite{brown2024modular}. In the present study, the RIS simulations use leaflet kinematics extracted from the IB results. Accordingly, the comparison is intended to assess how well a prescribed kinematics RIS formulation can reproduce hemodynamic quantities of interest relative to a fully coupled IB reference. The two workflows are compared in terms of bulk flow patterns, velocity fields, and transvalvular pressure gradient. In addition, we demonstrate that the RIS method can be combined with a monolithic ALE FSI formulation for an elastic arterial wall within svMultiPhysics, and we assess how wall compliance affects the transient hemodynamic response.

The remainder of the paper is organized as follows. Section~\ref{sec:methods} presents the mathematical formulations of the RIS and IB methods together with their implementations in the svMultiPhysics and IBAMR simulation frameworks. Section~\ref{sec:results} presents the numerical results for the trileaflet valve, bicuspid valve, and elastic wall RIS simulations. Section~\ref{sec:discussion} discusses the advantages, limitations, and complementary roles of the two methods. Section~\ref{sec:conclusion} summarizes the main conclusions.


\section{Heart valve modeling methods} \label{sec:methods}

This section presents the mathematical formulations of the RIS and IB methods employed in this study, followed by their implementations in the open-source simulation frameworks svMultiPhysics~\cite{zhu2022svfsi, svmultiphysics-code} and IBAMR~\cite{griffith2017ibamr, ibamr-code}. Both techniques enable the modeling of fluid--leaflet interaction on nonconforming meshes. The main distinction between them is how the valve leaflets are modeled and how forces are transferred between the fluid and the structure.

\subsection{Resistive immersed surface (RIS) method} \label{subsec:ris-method}

In the RIS method, the valve leaflets are represented as surface meshes whose opening and closing motions are known a priori. The valve motions can be obtained from medical imaging data, structural mechanics simulations, or other FSI analyses and be provided as input to the RIS heart valve simulation.

Let $\Omega$ denote the background fluid domain and $\Gamma_i$ be the surface of the $i$-th valve leaflet. The motion of the leaflet is described by a displacement field $\bm{d}_{\Gamma_i}$, from which the leaflet velocity is obtained as
\begin{align}
\bm{u}_{\Gamma_i} = \frac{\partial \bm{d}_{\Gamma_i}}{\partial t} \text{ .}\label{eq:ris_velocity}
\end{align}

The prescribed leaflet motion exerts a resistive force on the surrounding fluid through a penalty-type formulation
\begin{align}
\bm{f}_{\Gamma_i} = \frac{R_i}{\varepsilon_i}\, \delta_i\!\left(\varphi_i(\bm{x})\right) \left(\bm{u}-\bm{u}_{\Gamma_i}\right) \text{ ,} \label{eq:ris_resistive_force}
\end{align}
where $\bm{u}$ is the fluid velocity, $R_i$ is the resistance coefficient associated with the leaflet surface, and $\varepsilon_i$ represents the half-thickness of the immersed valve surface. The variable $\bm{x}$ denotes a spatial location in the fluid domain. $\varphi_i(\bm{x})$ is the signed distance from a fluid point $\bm{x}$ to the leaflet surface $\Gamma_i$, defined as
\begin{align}
\varphi_i(\bm{x}) =\begin{cases}
+ \|\bm{x}-\bm{y}^*\| \text{ ,} & (\bm{x}-\bm{y}^*)\cdot \bm{n}_i(\bm{y}^*) > 0 \text{ ,}\\
- \|\bm{x}-\bm{y}^*\| \text{ ,} & (\bm{x}-\bm{y}^*)\cdot \bm{n}_i(\bm{y}^*) < 0 \text{ ,}
\end{cases} \label{eq:ris_sdf}
\end{align}
where 
\[\bm{y}^* = \arg\min_{\bm{y}\in\Gamma_i} \|\bm{x}-\bm{y}\|\]
is the closest point on the surface $\Gamma_i$ to $\bm{x}$, and $\bm{n}_i$ is the outward-facing unit normal vector of the leaflet surface.

The resistive region is defined by the smoothed Dirac delta function $\delta_i$, whose support size is determined by the half-thickness of the valve. The smoothed Dirac delta function is given by
\begin{align}
\delta(y) = \begin{cases}
\frac{1}{2\varepsilon_i}\left(1+\cos\left(\frac{\pi y}{\varepsilon_i}\right)\right) \text{ ,} & \vert y \vert \le\varepsilon_i \text{ ,}\\
0 \text{ ,} & \vert y \vert > \varepsilon_i \text{ .}
\end{cases} \label{eq:ris_dirac_delta_function}
\end{align}
The incompressible Navier--Stokes equations augmented with RIS forces are
\begin{equation}
 \begin{aligned}
\rho\left(\frac{\partial \bm{u}}{\partial t}+(\bm{u}\cdot\nabla)\bm{u}\right)-\nabla\cdot\bm{\sigma}+\sum_{i=1}^{n}\bm{f}_{\Gamma_i} &=\bm{f}_b \quad \text{ in } \Omega  \text{ ,}\\
\nabla\cdot\bm{u} &=0  \quad \, \text{ in } \Omega \text{ ,} \label{eq:NS_ris}
\end{aligned}   
\end{equation}
where $\rho$ is the fluid density, $\bm{f}_b$ denotes body forces, and $\bm{\sigma}$ is the Cauchy stress tensor,
\begin{align}
\bm{\sigma} = -p\bm{I} + 2\mu\bm{\epsilon} \text{ ,} \qquad
\bm{\epsilon} = \frac{1}{2}\left(\nabla\bm{u} + \nabla\bm{u}^\mathrm{T}\right) \text{ ,} \label{eq:fluid_stress_strain}
\end{align}
with $p$ the pressure, $\mu$ the dynamic viscosity, and $\bm{\epsilon}$ the strain tensor.

When the fluid domain is solved within an ALE framework, the mesh velocity is denoted by $\hat{\bm{u}}$. The momentum equation then becomes
\begin{align}
\rho\left(\frac{\partial \bm{u}}{\partial t}+\left((\bm{u}-\hat{\bm{u}})\cdot\nabla\right)\bm{u}\right) - \nabla\cdot\bm{\sigma}+\sum_{i=1}^{n}\bm{f}_{\Gamma_i}=\bm{f}_b \quad \text{ in } \Omega \text{ .} \label{eq:NS_ris_ale}
\end{align}
Under the assumption that the smeared valve surfaces follow the motion of the fluid mesh, the RIS force term~\cite{zingaro2023modeling, africa2024lifex} is given by
\begin{align}
\bm{f}_{\Gamma_i} =\frac{R_i}{\varepsilon_i}\,\delta_i\!\left(\varphi_i(\bm{x})\right)\left(\bm{u}-\hat{\bm{u}}-\bm{u}_{\Gamma_i}\right) \text{ .} \label{eq:ris_resistive_force_ale}
\end{align}

Since the valve motion is prescribed, the RIS formulation in the present work should be interpreted as a reduced hemodynamic model conditioned on leaflet kinematics, rather than as a predictive model of valve mechanics. It can reproduce the flow response associated with a specified valve motion, but it cannot determine that motion from fluid loading and therefore cannot predict regurgitation, leaflet stresses, or sealing performance.

\subsection{Immersed boundary (IB) method} \label{subsec:ib-method}

In contrast to the RIS approach, the IB method models the two-way coupling between the fluid and the valve leaflets. The fluid motion influences the structural dynamics of the valve, and the resulting structural forces are applied back to the fluid.

In the IB formulation, the fluid velocity is denoted by $\bm{u}(\bm{x},t)$, defined on an Eulerian fluid grid at spatial location $\bm{x}$ and time $t$. The force exerted by the structure on the fluid is represented by $\bm{f}(\bm{x},t)$. The valve leaflets are described in a Lagrangian framework, where $\bm{X}(\bm{s},t)$ denotes the spatial position of a structural material point labeled by the Lagrangian coordinate $\bm{s}$ at time $t$, and $\bm{F}(\bm{s},t)$ represents the structural force density.

The fluid motion is governed by the incompressible Navier--Stokes equations. In the IB method, the structural force is transferred to the fluid grid through a force spreading operation using the Dirac delta function
\begin{align}
\bm{f}(\bm{x},t)=\int_{\Gamma}\bm{F}(\bm{s},t)\,\delta\!\left(\bm{x}-\bm{X}(\bm{s},t)\right)\,\mathrm{d}\bm{s} \text{ .} \label{eq:ib_force}
\end{align}
The structural force density is typically obtained from a constitutive model
\begin{align}
\bm{F}(\cdot,t)=\mathcal{F}\!\left(\bm{X}(\cdot,t)\right) \text{ ,}
\end{align}
where the omitted argument indicates that the force $\bm{F}$ may depend on the entire structure configuration $\bm{X}$. To enforce the no-slip condition between the structure and the fluid, the velocity of the immersed structure is obtained by interpolating the fluid velocity to the Lagrangian points
\begin{align}
\frac{\partial \bm{X}(\bm{s},t)}{\partial t}=\int_{\Omega}
\bm{u}(\bm{x},t)\,\delta\!\left(\bm{x}-\bm{X}(\bm{s},t)\right) \,\mathrm{d}\bm{x} \text{ .} \label{eq:ib_velocity}
\end{align}

The IB formulation therefore consists of four coupled components, including the incompressible Navier--Stokes equations for the fluid, the force spreading operator, the structural constitutive model, and the velocity interpolation operator. Together, these components constitute a two-way coupled FSI system for the immersed valve leaflets.

When the IB formulation is discretized, the Dirac delta function is replaced by a smoothed, approximate delta function $\delta_h$, where $h$ is the fluid mesh size. The smoothed delta function controls how structural forces are spread to the fluid grid and how fluid velocities are interpolated back. In practice, $\delta_h$ is constructed as a tensor product of one-dimensional functions with compact support. Common options include the 3-point and 4-point kernels~\cite{roma1999adaptive, peskin2002immersed}, which offer a good balance of smoothness, accuracy, and computational efficiency. Other kernels, e.g., 5-point and 6-point~\cite{bao2015gaussian, bale2021one}, as well as B-spline-based kernels~\cite{gruninger2025composite} offer smoother force distributions and improved translational invariance at the cost of slightly increased computational effort. In this study, we adopt the 5-point kernel, which is widely used in cardiovascular simulations and provides stable and accurate force spreading for aortic valve leaflets~\cite{bao2015gaussian}.

In summary, the RIS method represents the valve leaflets through a resistive force term associated with prescribed leaflet motion, leading to one-way coupling from the valve kinematics to the fluid flow. In contrast, the IB method explicitly resolves the two-way interaction between the fluid and the valve structure by coupling the structural dynamics to the fluid equations. Although both methods allow immersed valve leaflets to be represented on nonconforming meshes, the IB method provides a more physically complete description of valve mechanics.


\subsection{Open-source implementation of the RIS method in svMultiPhysics} \label{subsec:software-ris}

The RIS valve model is implemented in the parallel finite element multiphysics solver svMultiPhysics, which is designed for cardiovascular simulations and includes physics modules for hemodynamics, FSI, cardiac mechanics, and cardiac electrophysiology. svMultiPhysics is part of the open-source SimVascular project~\cite{updegrove2017simvascular}, which provides an integrated pipeline from medical image segmentation to patient-specific cardiovascular simulation and analysis. In this work, the existing svMultiPhysics framework is extended to incorporate RIS valve modeling.

\subsubsection{svMultiPhysics solver} \label{subsubsec:svmultiphysics-solver}

The svMultiPhysics solver employs the residual-based variational multiscale (VMS) method~\cite{bazilevs2007variational, liu2018unified} to discretize the incompressible flow equations. This formulation enables the use of equal-order interpolation by providing a consistent pressure stabilization effect through the fine-scale model.
  
In the VMS formulations, the solution is decomposed into resolved and unresolved scale components
\begin{align}
    \bm{u} = \bm{u}^h + \bm{u}' \qquad \text{and} \qquad p = p^h + p' \text{ ,}
\end{align}
with the unresolved scales approximated by
\begin{align}
    \bm{u}'\approx -\tau_M \bm{r}_M \qquad \text{and} \qquad p'\approx -\tau_C r_C \text{ ,}
\end{align}
where $\bm{r}_M$ and $r_C$ denote the momentum and continuity residuals of the Navier--Stokes equations, respectively.

By expressing the unresolved scales in terms of the governing equation residuals, the VMS method introduces stabilization terms that remain consistent with the original formulation while suppressing nonphysical oscillations. This makes the method particularly suitable for stable simulation of convection-dominated and high-Reynolds-number flows. The VMS formulation may be interpreted as a large-eddy simulation (LES)-type model with implicit subgrid-scale dissipation. Within the finite element framework of svMultiPhysics, the same formulation also supports conforming monolithically coupled FSI simulations of elastic arterial walls. When the RIS force term is included in this monolithic FSI system, the standard implicit coupling strategy, where the fluid--structure and mesh equations are iterated alternately until convergence, exhibits convergence difficulties. In this case, an explicit geometric coupling strategy~\cite{bucelli2022partitioned} is used, where the monolithic fluid--structure system is first solved to convergence and the mesh equation is then solved separately before advancing to the next time step.

svMultiPhysics uses the generalized-$\alpha$ method~\cite{Hul93, JanWhi99} for time integration. This second-order implicit scheme is widely used in computational fluid dynamics for transient simulations. The method evaluates the governing equations and enforces the residuals at intermediate time levels, thereby introducing controllable numerical dissipation of high-frequency modes while preserving the accuracy of the low-frequency physical phenomena. This property improves both stability and robustness for convection-dominated flows. Consequently, the generalized-$\alpha$ method enables accurate transient simulation while damping spurious numerical oscillations without significantly affecting the physical solution.

\subsubsection{RIS implementation in svMultiPhysics} \label{subsubsec:ris-implementation}

The RIS method is implemented by extending the fluid solver routines in the svMultiPhysics framework. The input to the RIS valve model consists of a set of leaflet surface meshes together with the corresponding nodal positions that prescribe the valve kinematics during the opening and closing phases. The nodal positions of the valve leaflets are denoted by $\bm{X}^{n_{\text{RIS}}}$, where the index $n_{\text{RIS}}$ runs from $1$ to $n_{\text{open}}$ during the opening phase and from $1$ to $n_{\text{close}}$ during the closing phase. The leaflet velocity is approximated using a backward difference rule
\begin{align}
    \bm{u}_{\Gamma}^{n_{\text{RIS}}} = \frac{\bm{X}^{n_{\text{RIS}}} - \bm{X}^{n_{\text{RIS}}-1}}{\Delta t} \text{ ,}
\end{align}
and $\bm{u}_{\Gamma}=\bm{0}$ when the valve is not moving. The resulting valve velocity field is then interpolated from the leaflet surface to the surrounding fluid elements.

To evaluate the resistive term in Eq.~\eqref{eq:ris_resistive_force}, a signed distance function (SDF) field is first constructed for the valve according to Eq.~\eqref{eq:ris_sdf}. The SDF $\varphi$ takes negative values on the upstream side. Based on the computed SDF field, the smoothed Dirac delta function is evaluated using Eq.~\eqref{eq:ris_dirac_delta_function}. The support of this smoothed Dirac delta function defines the resistive region associated with the valve, where $\varepsilon$ denotes the half-thickness. As noted in~\cite{fedele2017patient}, $\varepsilon$ is typically chosen to be approximately $1.5\,h$, where $h$ is the local fluid mesh size near the leaflet surface. This scaling produces a total resistive band roughly three fluid elements wide in the leaflet thickness direction so that the smoothed Dirac delta in Eq.~\eqref{eq:ris_dirac_delta_function} can be sampled at Gauss quadrature points of multiple adjacent elements, yielding a smooth and well-resolved resistive force distribution. For the unstructured tetrahedral meshes used in this work, $h$ is taken as a representative local element size in the leaflet region. For a user-specified resistance parameter $R$, the resistive force term in Eq.~\eqref{eq:ris_resistive_force} is evaluated at Gauss quadrature points of the fluid elements and assembled into the discretized equations, thereby contributing to both the residual vector and the tangent matrix of the discrete system.

Only the opening and closing motions are prescribed in the simulation, while the transition between these two phases is determined dynamically from the fluid solution. When the valve is in the closed state, the mean pressures on the upstream and downstream sides of the valve, denoted by $\bar{p}_{\text{up}}$ and $\bar{p}_{\text{down}}$, are are computed by averaging the pressure over the regions where the SDF satisfies $\varepsilon < \vert \varphi \vert \leq 5\, \varepsilon$ on the upstream and downstream sides, respectively. If $\bar{p}_{\text{up}} > \bar{p}_{\text{down}}$, the valve is assumed to begin opening. Once the valve reaches the fully open state, the averaged flow rate $\bar{Q}$ through the valve region is checked at each time step. The averaged flow rate $\bar{Q}$ is computed by integrating the velocity over the volume enclosed within the valve, defined as the region where $\varphi \leq -\varepsilon$ within the bounding box of the valve surface. If $\bar{Q} < 0$, the closing motion is activated. Thus, although the leaflet kinematics during opening and closing are prescribed, the switching between these two phases is determined by the fluid state.

The computational workflow of the RIS method for valve simulation in the svMultiPhysics framework is summarized in Algorithm~\ref{alg:ris_workflow}.

\begin{algorithm}[t]
\caption{Workflow of the RIS valve simulation in svMultiPhysics}
\label{alg:ris_workflow}
\begin{algorithmic}[1]
\State {\textbf{Input:} Fluid mesh, heart valve surface meshes, prescribed leaflet kinematics,

\quad resistance parameter $R$, half-thickness $\varepsilon$}
\For{each time step}
    \If{valve is opening or closing}
        \State Update leaflet position $\bm{X}^{n_{\text{RIS}}}$ and SDF $\varphi$
        \State Compute $\bm{u}_{\Gamma}^{n_{\text{RIS}}}=(\bm{X}^{n_{\text{RIS}}}-\bm{X}^{n_{\text{RIS}}-1})/\Delta t$
        \State Interpolate $\bm{u}_{\Gamma}$ to surrounding fluid elements
    \Else
        \State Set $\bm{u}_{\Gamma}=\bm{0}$
    \EndIf

    \State Evaluate the smoothed Dirac delta function and identify the resistive region
    \State Assemble the RIS resistive contribution in residual vectors and tangent matrices
    \State Solve the discrete system with Newton's iterations
    \If{valve is closed}
        \State Compute $\bar{p}_{\text{up}}$ and $\bar{p}_{\text{down}}$
        \If{$\bar{p}_{\text{up}} > \bar{p}_{\text{down}}$}
            \State Activate valve opening
        \EndIf
    \ElsIf{valve is open}
        \State Compute averaged flow rate $\bar{Q}$
        \If{$\bar{Q} < 0$}
            \State Activate valve closing
        \EndIf
    \EndIf
\EndFor
\end{algorithmic}
\end{algorithm}

\subsection{IBAMR for IB simulation} \label{subsec:software-ib}

The two-way coupled FSI simulations of blood flow and valve leaflet dynamics are performed using the open-source framework IBAMR~\cite{griffith2017ibamr, ibamr-code}, as performed in~\cite{kaiser2021design, kaiser2022controlled, kaiser2023comparison, kaiser2024simulation, kaiser2025simulation}. IBAMR implements the IB method, in which the fluid equations are solved on an Eulerian grid, while the immersed structure is represented on a Lagrangian mesh, as described in Section~\ref{subsec:ib-method}. In this framework, the incompressible Navier--Stokes equations are discretized on a staggered Cartesian grid using a finite difference formulation, and the valve leaflets are described by a Lagrangian solid mesh embedded within the nonconforming fluid domain. The coupling between the fluid and the structure is achieved through the IB force-spreading~\eqref{eq:ib_force} and velocity-interpolation operators~\eqref{eq:ib_velocity} through a smoothed delta kernel. Because the smoothed delta kernel has finite support over several grid cells, the immersed structure has an effective numerical thickness that scales with the local Eulerian mesh spacing.

In IBAMR, the Navier--Stokes convective term is advanced using an Adams--Bashforth scheme, the diffusive term is treated implicitly, and the immersed boundary coupling terms are advanced using an explicit midpoint Runge--Kutta method. The fluid domain is discretized on a block-structured hierarchical Cartesian mesh, with optional adaptive mesh refinement provided through the SAMRAI library~\cite{hornung2002managing}.

\subsection{Comparative simulation setup} \label{subsec:simulation-setup}

\begin{figure}[!htb]
    \centering
    \hspace{\fill}
    \begin{subfigure}[t]{0.48\textwidth}
        \centering
        \includegraphics[width=\textwidth]{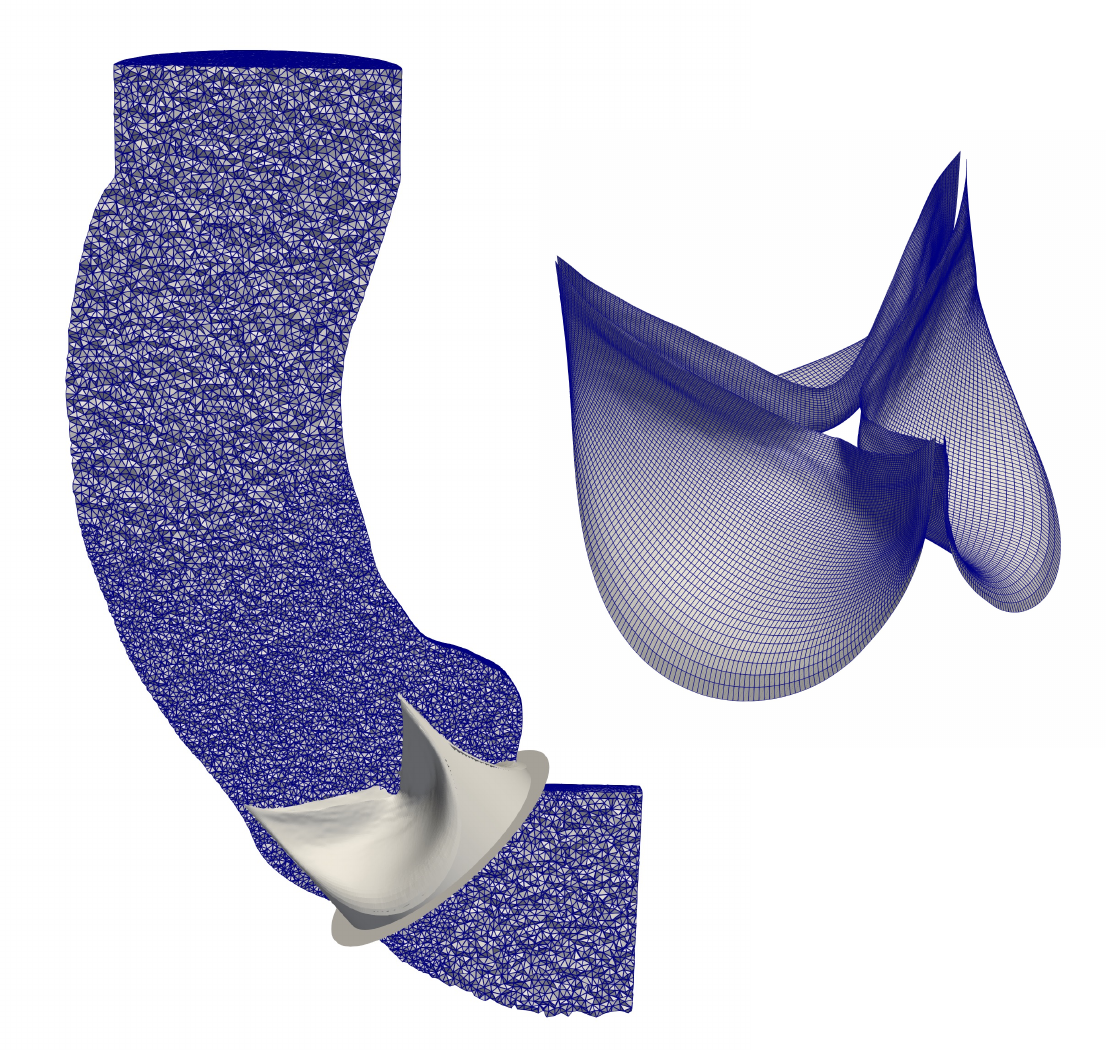}
        \caption{}
        \label{subfig:ris-trileaflet-valve-mesh}
    \end{subfigure}
    \hspace{\fill}
    \begin{subfigure}[t]{0.48\textwidth}
        \centering
        \includegraphics[width=\textwidth]{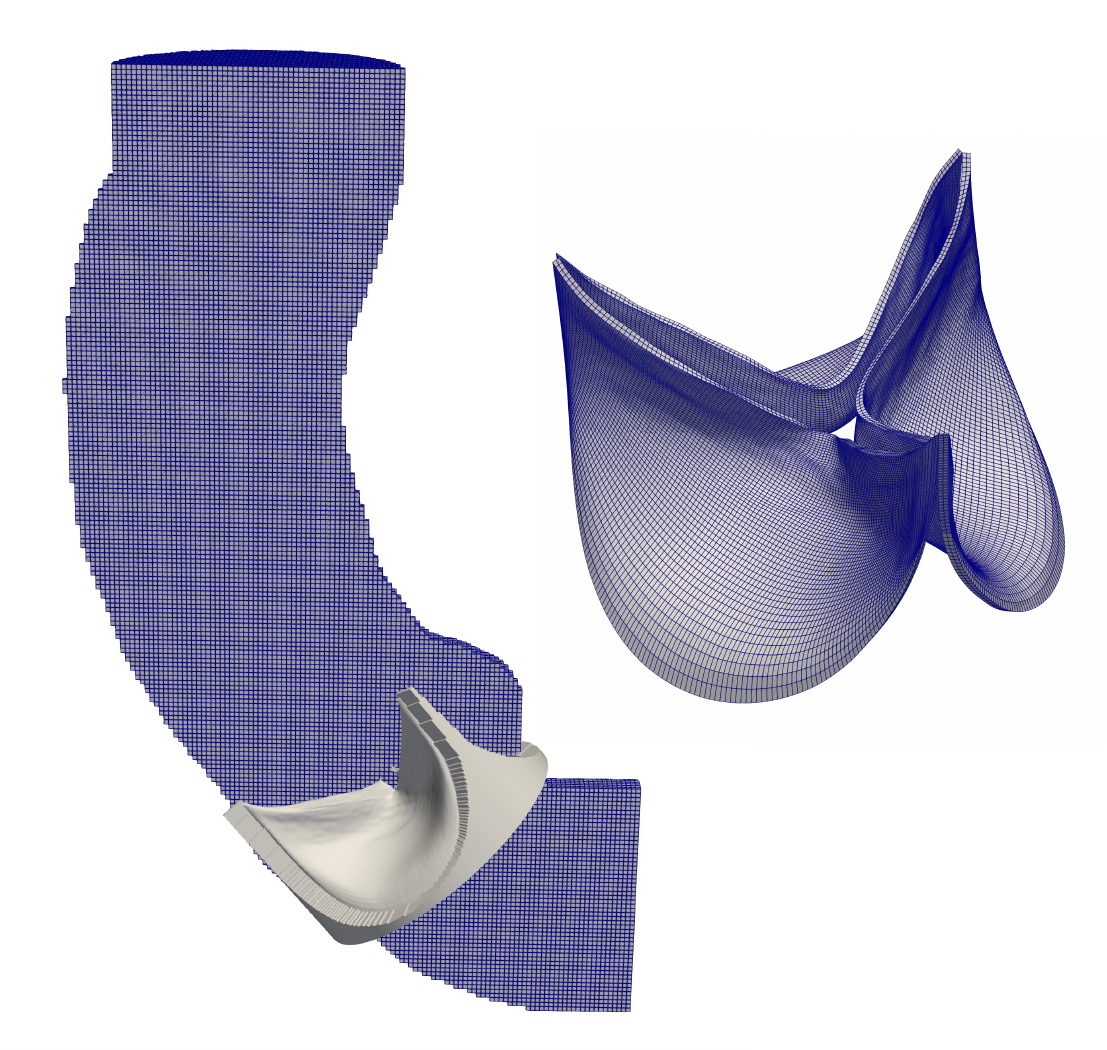}
        \caption{}
        \label{subfig:ib-trileaflet-valve-mesh}
    \end{subfigure}
    \hspace{\fill}
    \caption{(a) Finite element mesh of the aorta and surface mesh of the aortic valve used in the RIS simulation, and its location within the fluid domain. (b) Structured grid mesh of the aorta and solid mesh of the aortic valve used in the IB simulation, and its position relative to the fluid mesh.}
    \label{fig:arota-trileaflet-valve-mesh}
\end{figure}

The aortic valve hemodynamics are first simulated using the IB method on an Eulerian mesh with a grid size of 0.5~mm, which serves as the baseline in this comparative study. The IB simulation is performed in the IBAMR framework, described above. The structured grid used in the IB simulation has 192, 96, and 272 points in the $x$, $y$, and $z$ directions, respectively, for the full computational domain, while Figure~\ref{subfig:ib-trileaflet-valve-mesh} shows the aortic region after cropping the surrounding domain. The IB simulation serves as a fully coupled reference from which leaflet kinematics are extracted for the RIS simulation. Accordingly, the comparison is designed to assess the hemodynamic fidelity of RIS under prescribed leaflet motion rather than its ability to independently predict valve deformation. The corresponding unstructured tetrahedral mesh used in the RIS simulation for the aorta is shown in Figure~\ref{subfig:ris-trileaflet-valve-mesh}. The mesh size is targeted to edge length 1~mm in the bulk domain and is locally refined to 0.5~mm near the aortic valve region, giving a total of 1,736,056~degrees of freedom (DoFs).

\section{Numerical results}\label{sec:results}

This section presents the numerical results for three test cases. Sections~\ref{subsec:aortic-valve-comparison} and~\ref{subsec:bicuspid-valve-comparison} compare the RIS and IB methods on a trileaflet and a bicuspid aortic valve, respectively. Section~\ref{subsec:aortic-valve-fsi} demonstrates an extension in which the RIS valve is coupled to an elastic arterial wall through a monolithic ALE FSI formulation in svMultiPhysics. For each case, the results are visualized and compared, and the differences between the two workflows are analyzed.

\subsection{Aortic valve simulation comparison} \label{subsec:aortic-valve-comparison}

For the IB simulation, the valve leaflets are modeled by a Lagrangian mesh with nodal spacing of 0.22~mm, as shown in Figure~\ref{subfig:ib-trileaflet-valve-mesh}, and use a fully fiber-based, anisotropic material model. The fluid density is set to 1.0~$\mathrm{g}/\mathrm{cm^3}$, and a Newtonian viscosity model with viscosity 0.04~$\mathrm{g}/(\mathrm{cm}\cdot\mathrm{s})$ is used for both methods. The time step size for the IB simulation is $\Delta t = 6\times10^{-6}$~s. In the two-way coupling between the leaflets and the fluid, both force spreading from the structure to the fluid and velocity interpolation from the fluid to the leaflet mesh employ the IB5 kernel~\cite{bao2015gaussian}, which has a radius of 1.25~mm, which is equal to 2.5 times the fluid mesh width. 

Following the discussion in~\cite{fedele2017patient}, the valve half-thickness parameter $\varepsilon$ in the RIS method is chosen as 1.5 times the local element size near the valve surface. Since the local mesh size around the valve is 0.5~mm, $\varepsilon$ is set to 0.75~mm, corresponding to a total resistive region thickness of 1.5~mm. This scale is comparable to the interaction region thickness used in the IB simulation. In the RIS simulation, the resistance is set to $10^5~\mathrm{g}/(\mathrm{cm}\cdot \mathrm{s})$ to strongly suppress through surface flow.

\begin{figure}[!htb]
    \centering
    \hspace{\fill}
    \begin{subfigure}[t]{0.48\textwidth}
        \centering
        \includegraphics[width=\textwidth]{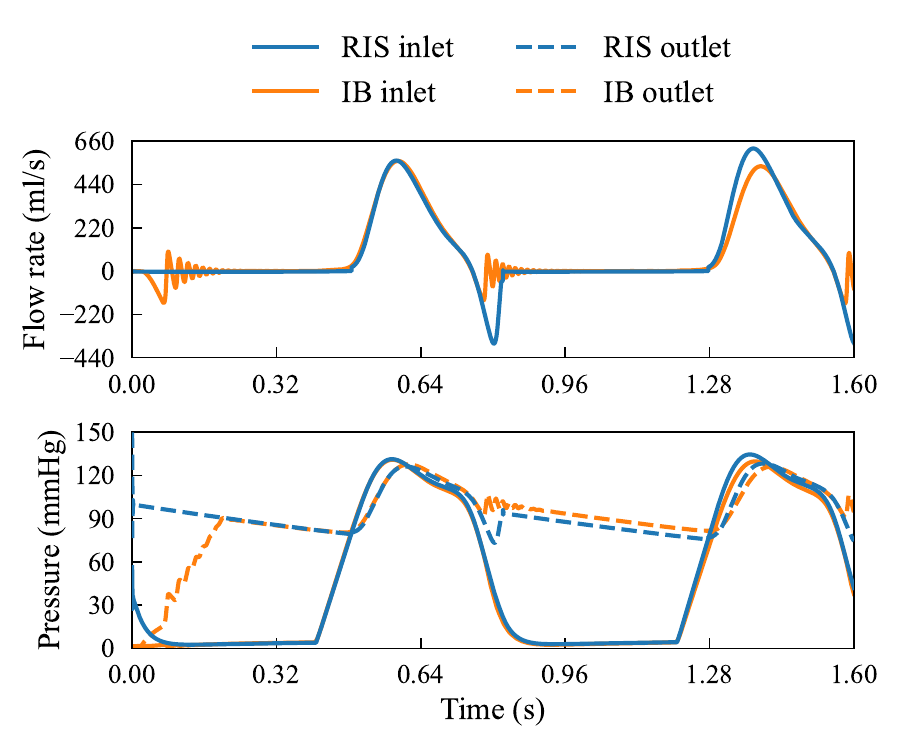}
        \caption{}
        \label{subfig:trileaflet-cfd-flowrate-pressure}
    \end{subfigure}
    \hspace{\fill}
    \begin{subfigure}[t]{0.48\textwidth}
        \centering
        \includegraphics[width=\textwidth]{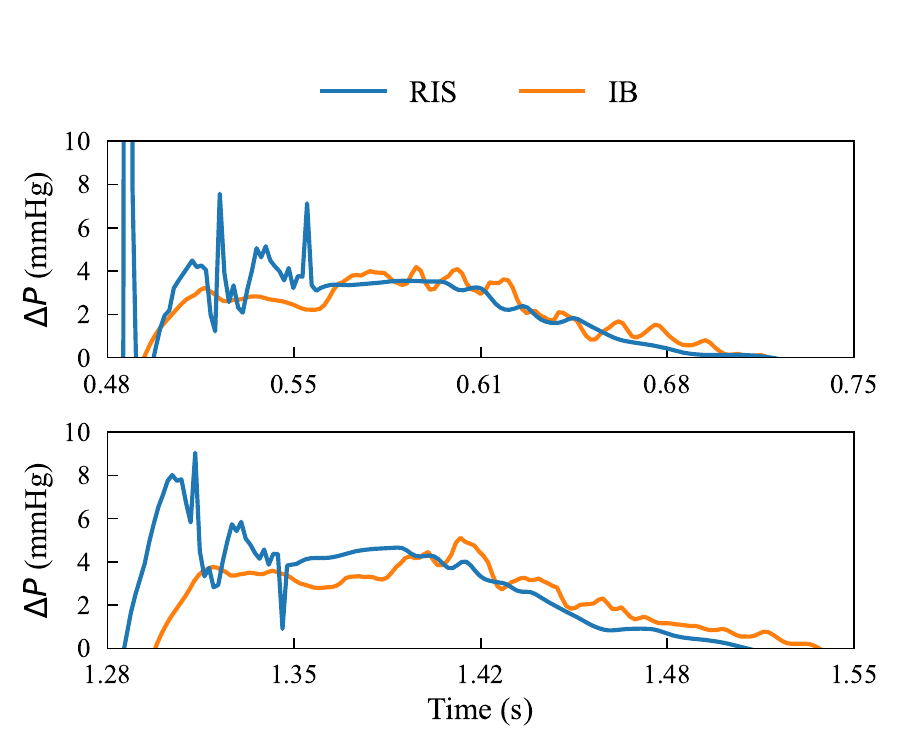}
        \caption{}
        \label{subfig:trileaflet-cfd-pressure-drop}
    \end{subfigure}
    \hspace{\fill}
    \caption{(a) Comparison of the flow rate and boundary pressure in the aortic valve simulations using the RIS and IB methods. (b) Comparison of the transvalvular pressure gradient $\Delta P$ predicted by the RIS and IB methods for the first and second cardiac cycles.}
    \label{fig:trileaflet-cfd-pressure-data}
\end{figure}

In both simulations, lumped-parameter models are used for the inlet and outlet boundary conditions of the aorta. A left ventricular chamber model is imposed at the aortic inlet to represent the pumping function of the heart~\cite{kim2009coupling}, while a Resistance--Compliance--Resistance (RCR) boundary condition is applied at the aortic outlet to represent the downstream vasculature. For the RIS simulation, these boundary conditions are provided through the coupling of svZeroDSolver~\cite{svzerodsolver-code, brown2024modular} with svMultiPhysics. In the IB simulation, the boundary conditions are implemented with explicit, partitioned coupling to the fluid solver. The RCR boundary condition is advanced via a backward Euler discretization and the ventricle model with an explicit algebraic update, both using the flow rate from the fluid solver explicitly. For the aortic wall treatment, the RIS simulation enforces a no-slip boundary condition on the wall. In contrast, the IB simulation models the aortic wall as an immersed structure held in place by ``tether points", stiff linear springs that maintain an approximately constant position. The boundary pressure and flow rate profiles from the IB simulation are shown in Figure~\ref{subfig:trileaflet-cfd-flowrate-pressure}. 

\begin{figure}[!htb]\centering
    \includegraphics[width=0.96\textwidth]{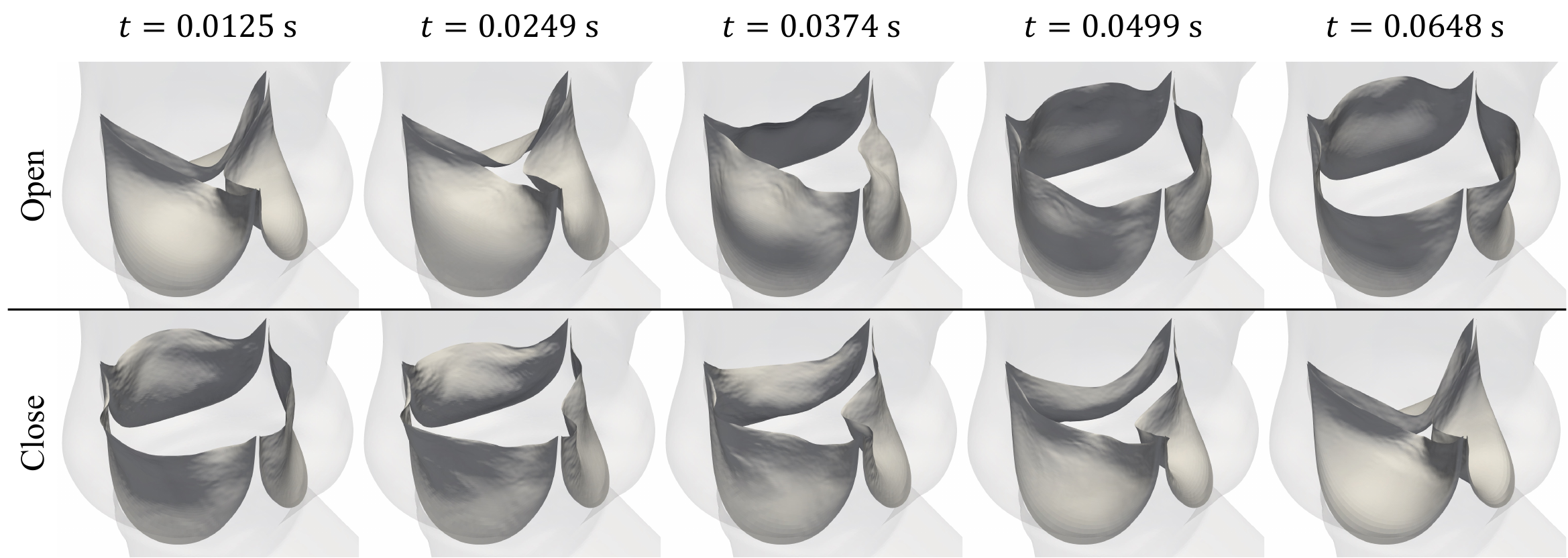}
    \caption{Snapshots of the prescribed opening and closing motions of the aortic valve used in the RIS simulation, where the leaflet kinematics are extracted from the IB simulation.}
    \label{fig:ris-trileaflet-motion}
\end{figure}

To enable a direct comparison, the valve kinematics $X_{\Gamma}$ used in the RIS simulation are extracted from the IB results. Specifically, the mid-surfaces of the IB solid leaflet meshes are extracted and used as the valve surface meshes in the RIS simulation, as shown in Figure~\ref{subfig:ris-trileaflet-valve-mesh}. For the opening phase, the extracted motion corresponds to the initial opening of the valve during the first cardiac cycle, from a nearly closed state to a fully open state, over a total duration of 0.0648~s. Similarly, a sequence of leaflet configurations during closure is extracted from the IB simulation, spanning the motion from fully open to fully closed, and is prescribed over the same 0.0648~s interval. The closing motion is still derived from the IB results, but its timing is condensed relative to the gradual diastolic settling observed in the IB simulation, where the leaflets close over a longer interval with smoother transitions. This condensation is a deliberate modeling choice since prescribing the full IB diastolic motion would require a large number of displacement steps as kinematic input. Furthermore, combined with the closed motion criterion in Section~\ref{subsec:software-ris}, it would result in a prolonged partially open state and more pronounced backflow. As a consequence, the RIS closing transient is faster than that in the IB simulation, and differences during valve closure are expected. Representative snapshots of the extracted opening and closing motions are shown in Figure~\ref{fig:ris-trileaflet-motion}. These extracted opening and closing sequences provide a controlled kinematic input for RIS. Because they are derived from the IB simulation, the resulting comparison is partially conditioned on the IB reference. The RIS simulation uses a time step size of $\Delta t = 4.155\times10^{-4}$~s, resulting in 157 time steps for the prescribed opening and closing motion. In both simulations, a scaffold geometry is included to prevent leakage between the valve and the aortic wall. The scaffold is modeled using solid elements in the IB simulation and as a surface mesh in the RIS simulation, and it remains fixed throughout the simulation.

\begin{figure}[!htb]\centering
    \includegraphics[width=0.96\textwidth]{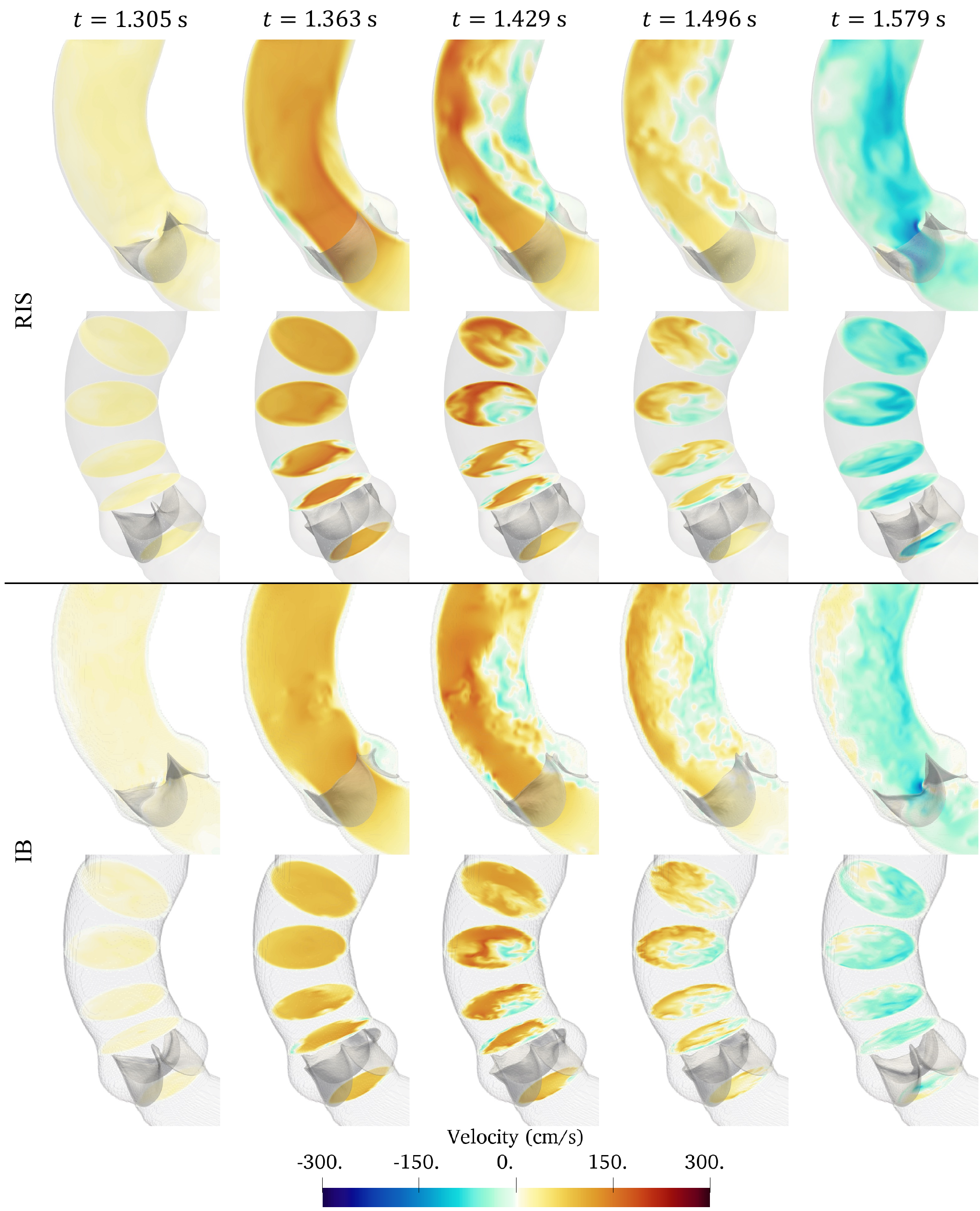}
    \caption{Comparison of the vertical velocity component and bulk flow structures in the aortic valve simulations obtained with the RIS and IB methods during the second cardiac cycle. The slices shown in the second row are normal to the artery axis, and those located immediately below and above the valve are used to calculate the transvalvular pressure gradient.}
    \label{fig:trileaflet-cfd}
\end{figure}

With the aorta and valve meshes, the prescribed valve motion in the RIS simulation, and the same lumped-parameter boundary conditions, simulations are performed for the RIS method. Representative snapshots of the fluid velocity field in the second cardiac cycle are shown in Figure~\ref{fig:trileaflet-cfd}. The two simulations show similar large-scale systolic flow patterns and comparable velocity magnitudes during valve opening. The main discrepancies arise during valve closure and in the second cycle, where the prescribed RIS kinematics produce increased reversed flow and altered coupling with the chamber and outlet RCR models. Although the prescribed RIS closing motion starts from fully open to fully closed within 0.0648~s, the sweep is not initiated until the flow state criterion in Section~\ref{subsec:software-ris} detects reversed flow at the valve. 

In the IB simulation, by contrast, the leaflets respond continuously to fluid loading and begin moving toward closure as soon as an adverse transvalvular pressure gradient develops, during late systolic deceleration and before flow has actually reversed. So IB closure effectively starts earlier in the cycle, even though the leaflet motion itself develops more gradually. During the interval between the earlier IB closure and the point at which the RIS closure criterion is reached, the RIS valve remains fully open, allowing reverse flow to pass through and accumulate as backflow. Consequently, despite the shorter prescribed RIS closing motion, the RIS simulation exhibits more pronounced backflow and a slight delay in closure timing relative to IB. Moreover, because the leaflet motion in RIS is prescribed rather than determined through two-way FSI, the closure cannot adapt dynamically to the instantaneous transvalvular pressure difference. In contrast, the IB simulation naturally couples leaflet motion and fluid loading, which improves the timing and effectiveness of closure. This difference is also evident in the flow rate curves shown in Figure~\ref{subfig:trileaflet-cfd-flowrate-pressure}.

For the first cardiac cycle, the flow rates predicted by the two methods agree well. The stroke volume is 78.80~ml for the RIS simulation and 81.34~ml for the IB simulation. In the second cardiac cycle, the RIS and IB stroke volumes are 88.62~ml and 77.08~ml, respectively. The larger stroke volume in the RIS simulation during the second cycle is associated with the larger backflow near the end of the first cycle, which alters the chamber dynamics and leads to a higher inlet pressure in the subsequent cycle: 134.52~mmHg for RIS versus 129.69~mmHg for IB, as shown in Figure~\ref{subfig:trileaflet-cfd-flowrate-pressure}. Furthermore, the RIS method predicts mean transvalvular pressure gradients of similar overall magnitude: 2.56~mmHg versus 2.23~mmHg for the first cardiac cycle, and 3.13~mmHg versus 2.47~mmHg for the second cycle, for RIS and IB, respectively. The pressure gradient is computed as the difference between the average pressures on two slices normal to the artery axis, located immediately below and above the valve, as shown in Figure~\ref{fig:trileaflet-cfd}. The pressure gradient curve in the RIS simulation is more oscillatory and contains sharper changes, as shown in Figure~\ref{subfig:trileaflet-cfd-pressure-drop}. This behavior is associated with the combined effects of the rigid aortic wall, the prescribed leaflet motion, and the absence of structural compliance in the leaflet dynamics.

\begin{figure}[!htb]
    \centering
    \hspace{\fill}
    \begin{subfigure}[t]{0.48\textwidth}
        \centering
        \includegraphics[width=\textwidth]{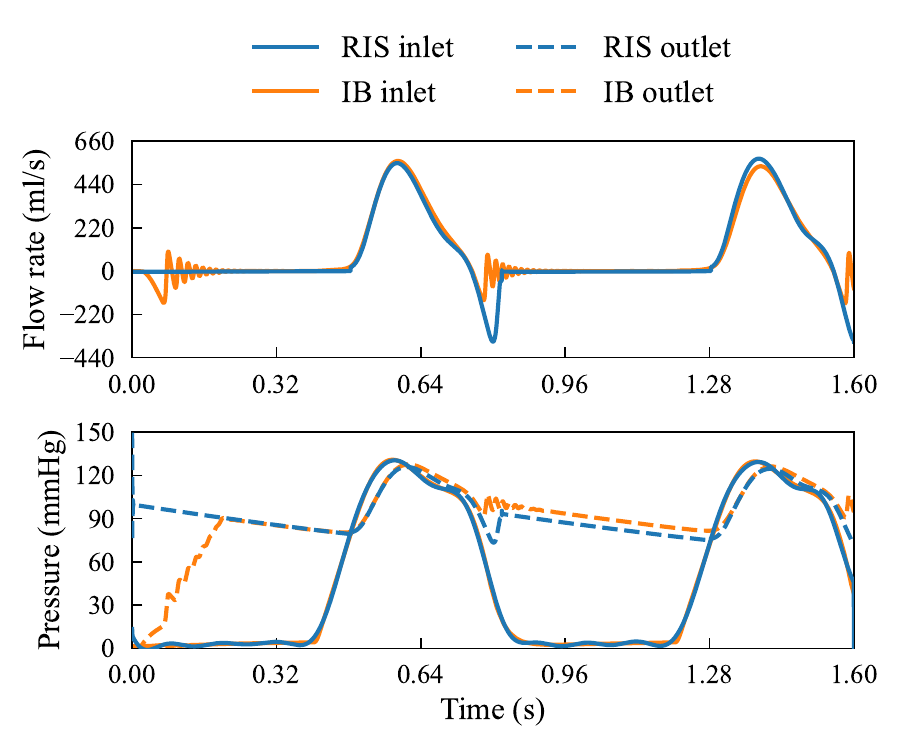}
        \caption{}
        \label{subfig:trileaflet-cfd-flowrate-pressure-pbc}
    \end{subfigure}
    \hspace{\fill}
    \begin{subfigure}[t]{0.48\textwidth}
        \centering
        \includegraphics[width=\textwidth]{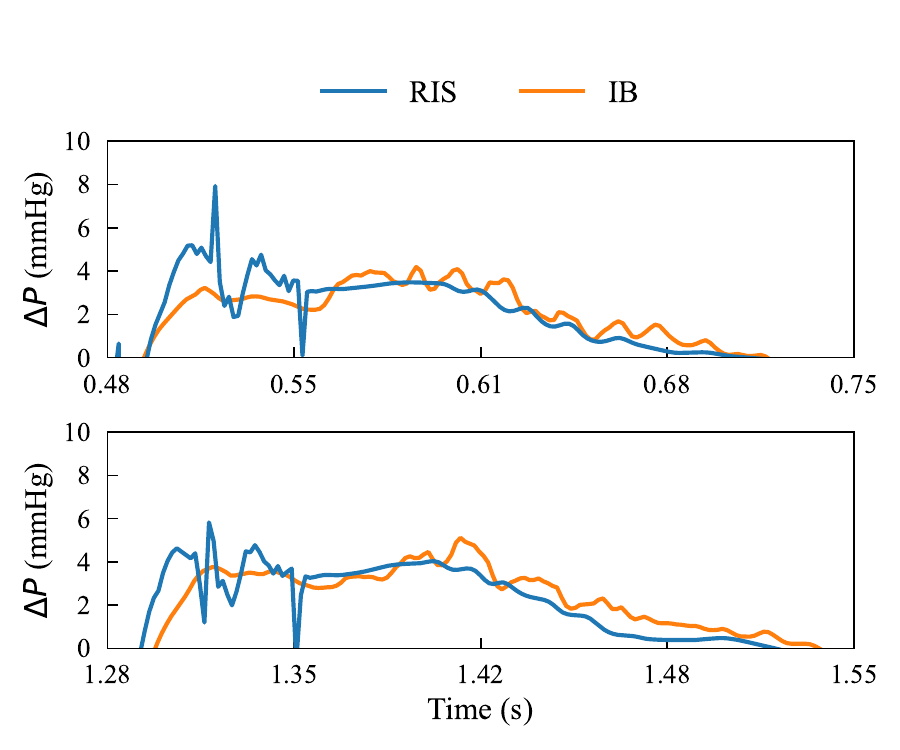}
        \caption{}
        \label{subfig:trileaflet-cfd-pressure-drop-pbc}
    \end{subfigure}
    \hspace{\fill}
    \caption{(a) Comparison of the flow rate and boundary pressure in the aortic valve simulations using the RIS and IB methods with prescribed inlet pressure applied in the RIS simulation. (b) Comparison of the transvalvular pressure gradient $\Delta P$ predicted by the RIS and IB methods for the first and second cardiac cycles with prescribed inlet pressure applied in the RIS simulation.}
    \label{fig:trileaflet-cfd-pressure-data-pbc}
\end{figure}

To further isolate the source of the second cycle discrepancy identified above, an additional test is performed in which the inlet boundary condition from the left ventricular chamber model is replaced by a prescribed pressure Neumann boundary condition extracted from the IB simulation, while the outlet boundary retains the same RCR condition. The resulting flow rates and boundary pressures are shown in Figure~\ref{subfig:trileaflet-cfd-flowrate-pressure-pbc}. By construction, the inlet pressures of the two simulations are now identical, thereby eliminating the chamber model response that amplifies the RIS backflow into a higher inlet driving pressure in the second cycle.

With the inlet pressure controlled, the forward flow rates in both cardiac cycles agree closely between the RIS and IB simulations, as shown in the top panel of Figure~\ref{subfig:trileaflet-cfd-flowrate-pressure-pbc}. This confirms that the substantial second cycle stroke volume difference observed in the original comparison is driven primarily by boundary condition feedback between the excess RIS backflow and the chamber model, rather than by the RIS valve formulation alone. The backflow spike at valve closure, which is characteristic of the prescribed RIS kinematics, persists in this test. However, without the chamber model to amplify its effect on the subsequent inlet pressure, its impact on the overall flow rates is greatly reduced. The outlet pressure, however, evolves differently between the two methods. Because the RCR boundary condition includes a compliant capacitor element, it integrates the net flow history over time. As a result, the larger backflow in the RIS simulation during valve closure charges the RCR capacitor to a different pressure state than in the IB simulation. This produces the visible offset in the outlet pressure curves shown in the lower panel of Figure~\ref{subfig:trileaflet-cfd-flowrate-pressure-pbc}. Consequently, the outlet pressure in the RIS simulation remains lower during the diastolic phase of the second cycle, which in turn produces a difference in the flow rate during that cycle.

The transvalvular pressure gradient for this prescribed inlet pressure test is shown in Figure~\ref{subfig:trileaflet-cfd-pressure-drop-pbc}. With the inlet pressure controlled, the mean pressure gradient between the two methods shows very good agreement: 2.34~mmHg versus 2.23~mmHg in the first cycle, and 2.44~mmHg versus 2.47~mmHg in the second cycle, for RIS and IB methods. The RIS pressure gradient curve remains more oscillatory than the IB result, and it rises more sharply at valve opening. Since the inlet condition has been controlled, these sharper oscillations can now be attributed primarily to the prescribed leaflet kinematics rather than to boundary condition effects. This test indicates that the main disagreement in second cycle stroke volume is largely due to closure-related reverse flow interacting with the coupled lumped-parameter boundary conditions. Accordingly, agreement between RIS and IB is stronger for direct flow and pressure gradient responses under controlled driving conditions than for fully coupled cycle-to-cycle dynamics.

Meanwhile, in Figure~\ref{fig:trileaflet-cfd}, the IB simulation retains more small-scale flow structures than the RIS simulation. One reason is that the svMultiPhysics simulation employs VMS stabilization, which introduces subgrid-scale dissipation and smoothes unresolved flow structures. In addition, the implicit generalized-$\alpha$ time integration used in svMultiPhysics is not constrained by the Courant-Friedrichs-Lewy (CFL) condition and allows significantly larger time step sizes than the explicit time integration used in IBAMR, providing improved computational efficiency. This large time step coarsens the temporal resolution and adds damping of high-frequency flow features. Another important factor is the leaflet motion itself. In the IB simulation, the leaflets continue to interact with the surrounding fluid through two-way coupling and show small-scale fluttering and swaying even after the valve has reached the fully open configuration, whereas in the RIS simulation, the leaflets remain perfectly stationary once the prescribed fully open state is reached and before the closing motion is initiated. This difference can also contribute to the more pronounced flow fluctuations observed in the IB results. 

\begin{figure}[!htb]\centering
    \includegraphics[width=0.96\textwidth]{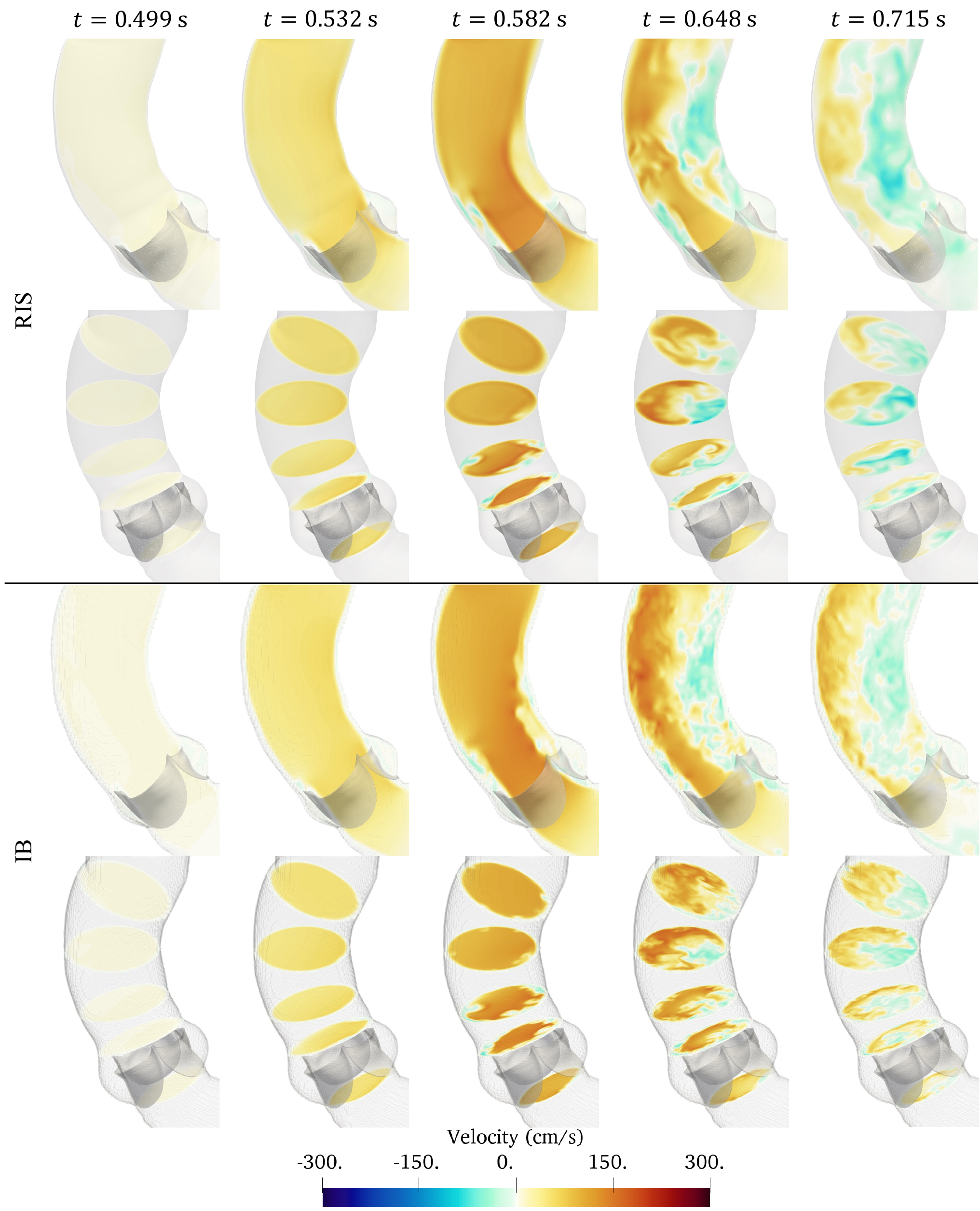}
    \caption{Comparison of the aortic flow velocity field in the vertical direction and bulk flow structures obtained with the RIS and IB methods when the aortic valve is fixed in the fully open position throughout the cardiac cycle.}
    \label{fig:trileaflet-cfd-fixed}
\end{figure}

To further examine the effect of leaflet flutter versus the numerical method, an additional test is performed in which the valve leaflets remain in the fully open position throughout the entire cardiac cycle in both simulations. Representative snapshots are shown in Figure~\ref{fig:trileaflet-cfd-fixed}. The comparison illustrates that svMultiPhysics and IBAMR produce similar large-scale flow patterns in this case, although the svMultiPhysics results still appear smoother because of the dissipation introduced by VMS stabilization and generalized-$\alpha$ time integration. The fixed-valve test helps separate these two effects. Even when the leaflets are identical and stationary in both simulations, the svMultiPhysics result is visibly smoother than the IBAMR result. This indicates that the smoother flow field in the RIS result partly comes from the implicit filtering in the VMS-based formulation and the larger time step enabled by the generalized-$\alpha$ time integration used in svMultiPhysics, both of which are expected features of the numerical scheme. The additional small-scale structures observed in the IB moving-valve case, which cannot be fully explained by the fixed-valve test, come from the leaflet motion itself. In IB, the leaflets flutter and sway under the unsteady fluid loading, and this small but continuous motion sheds vortices into the downstream flow. The RIS simulation cannot reproduce this effect because the leaflets remain stationary between the prescribed opening and closing motions.

\subsection{Bicuspid valve simulation comparison} \label{subsec:bicuspid-valve-comparison}

\begin{figure}[!htb]
    \centering
    \hspace{\fill}
    \begin{subfigure}[t]{0.48\textwidth}
        \centering
        \includegraphics[width=\textwidth]{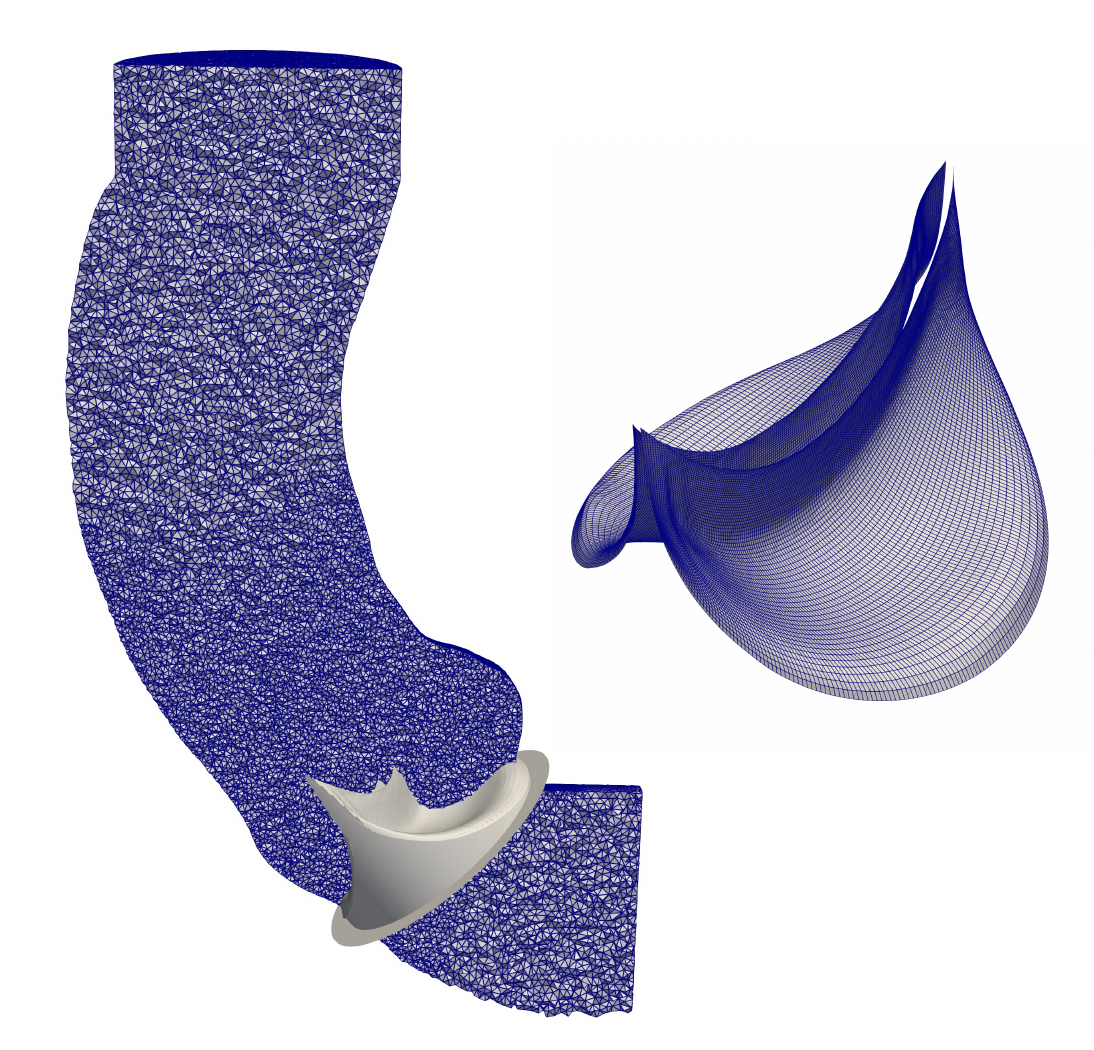}
        \caption{}
        \label{subfig:ris-bicuspid-valve-mesh}
    \end{subfigure}
    \hspace{\fill}
    \begin{subfigure}[t]{0.48\textwidth}
        \centering
        \includegraphics[width=\textwidth]{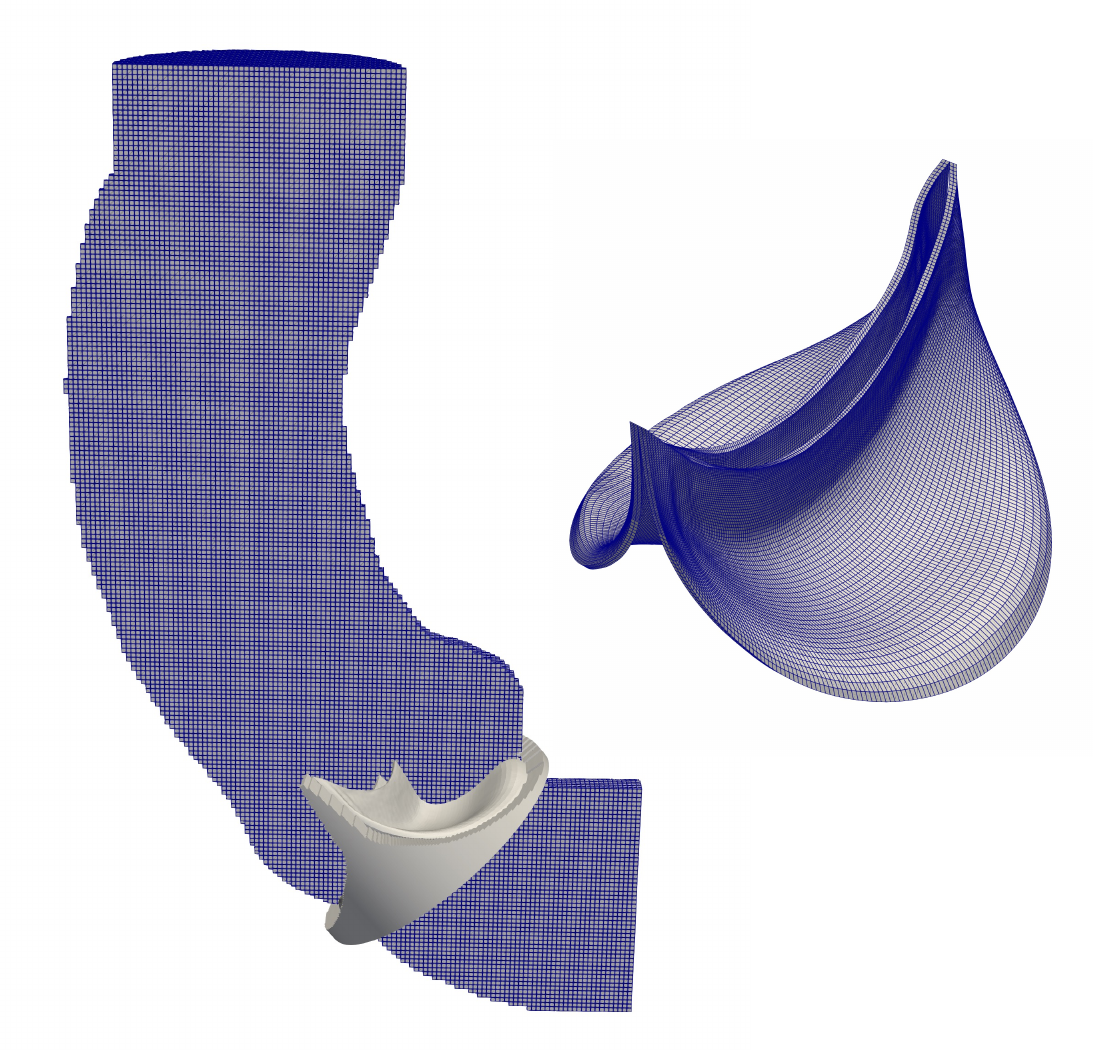}
        \caption{}
        \label{subfig:ib-bicuspid-valve-mesh}
    \end{subfigure}
    \hspace{\fill}
    \caption{(a) Finite mesh of the aorta and surface mesh of the bicuspid valve used in the RIS simulation, and its location within the fluid domain. (b) Structured grid mesh of the aorta and solid mesh of the bicuspid valve used in the IB simulation, and its position relative to the fluid mesh.}
    \label{fig:arota-bicuspid-valve-mesh}
\end{figure}

In this section, a bicuspid valve is used to further assess the RIS simulation by comparing its flow field with the IB results. This bicuspid model was originally constructed to study bicuspidization repair and is symmetric with no raphe~\cite{kaiser2024simulation}. To further test the pressure gradient predicted by the RIS method, we use a geometry that exhibited moderate stenosis in previous work. The free-edge length to annulus diameter ratio is 1.2, which showed a decreased effective orifice area and an elevated transvalvular pressure gradient relative to the optimal proposed geometry. The aortic meshes for the RIS and IB simulations are the same as those used in Section~\ref{subsec:aortic-valve-comparison}. As in the trileaflet case, an IB simulation is first performed as the baseline solution. The bicuspid valve solid mesh used in the IB simulation, together with the scaffold and its relative position within the aortic mesh, is shown in Figure~\ref{subfig:ib-bicuspid-valve-mesh}. The extracted mid-surface mesh used in the RIS simulation is shown in Figure~\ref{subfig:ris-bicuspid-valve-mesh}. The lumped-parameter boundary conditions are kept the same as in the trileaflet simulation. The purpose is to examine whether the RIS method can reproduce the transvalvular pressure gradient, which is a clinically relevant quantity, and whether it can capture bulk flow patterns similar to those obtained from the fully coupled IB simulation when the valve showed stenosis and produced an elevated gradient.

\begin{figure}[!htb]\centering
    \includegraphics[width=0.96\textwidth]{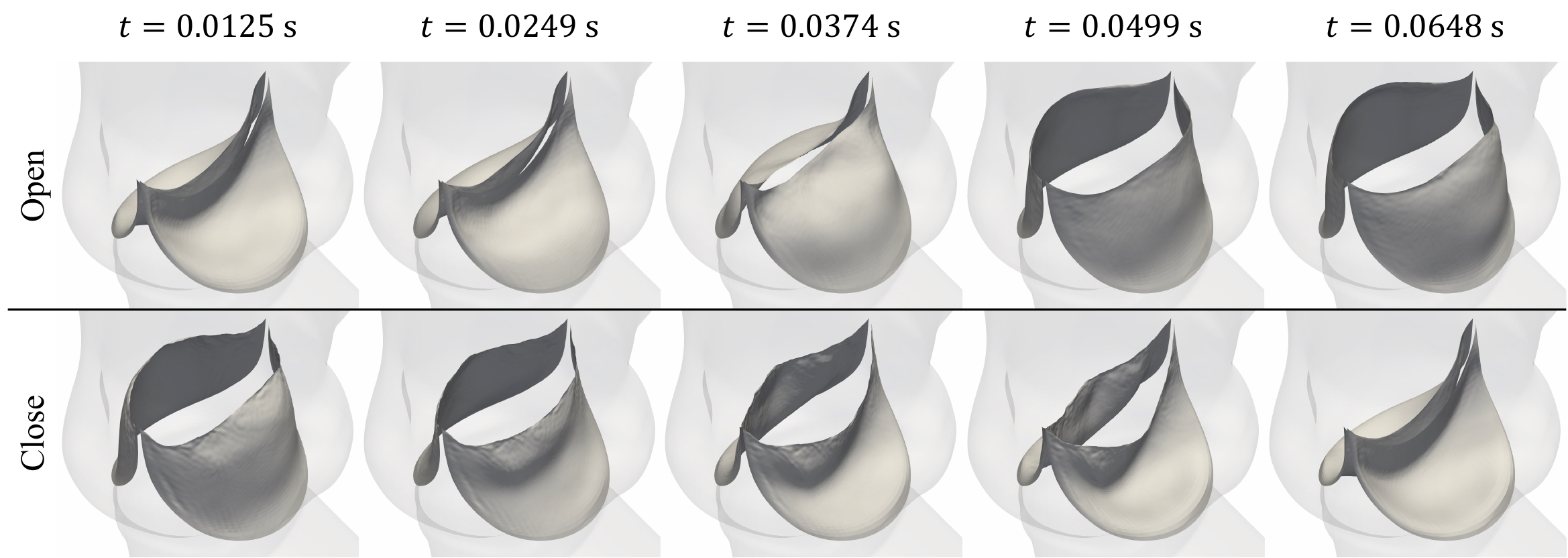}
    \caption{Snapshots of the prescribed opening and closing motions of the bicuspid valve used in the RIS simulation, where the leaflet kinematics are extracted from the IB simulation.}
    \label{fig:ris-bicuspid-motion}
\end{figure}

As in the previous case, a sequence of leaflet configurations is extracted from the IB simulation and prescribed in the RIS simulation, with a total duration of 0.0648~s for both the opening and closing motions. Representative snapshots of the RIS bicuspid valve motion are shown in Figure~\ref{fig:ris-bicuspid-motion}. The RIS half-thickness and resistance are also set to 0.75~mm and $10^5~\mathrm{g}/(\mathrm{cm}\cdot \mathrm{s})$, respectively.

\begin{figure}[!htb]\centering
    \includegraphics[width=0.96\textwidth]{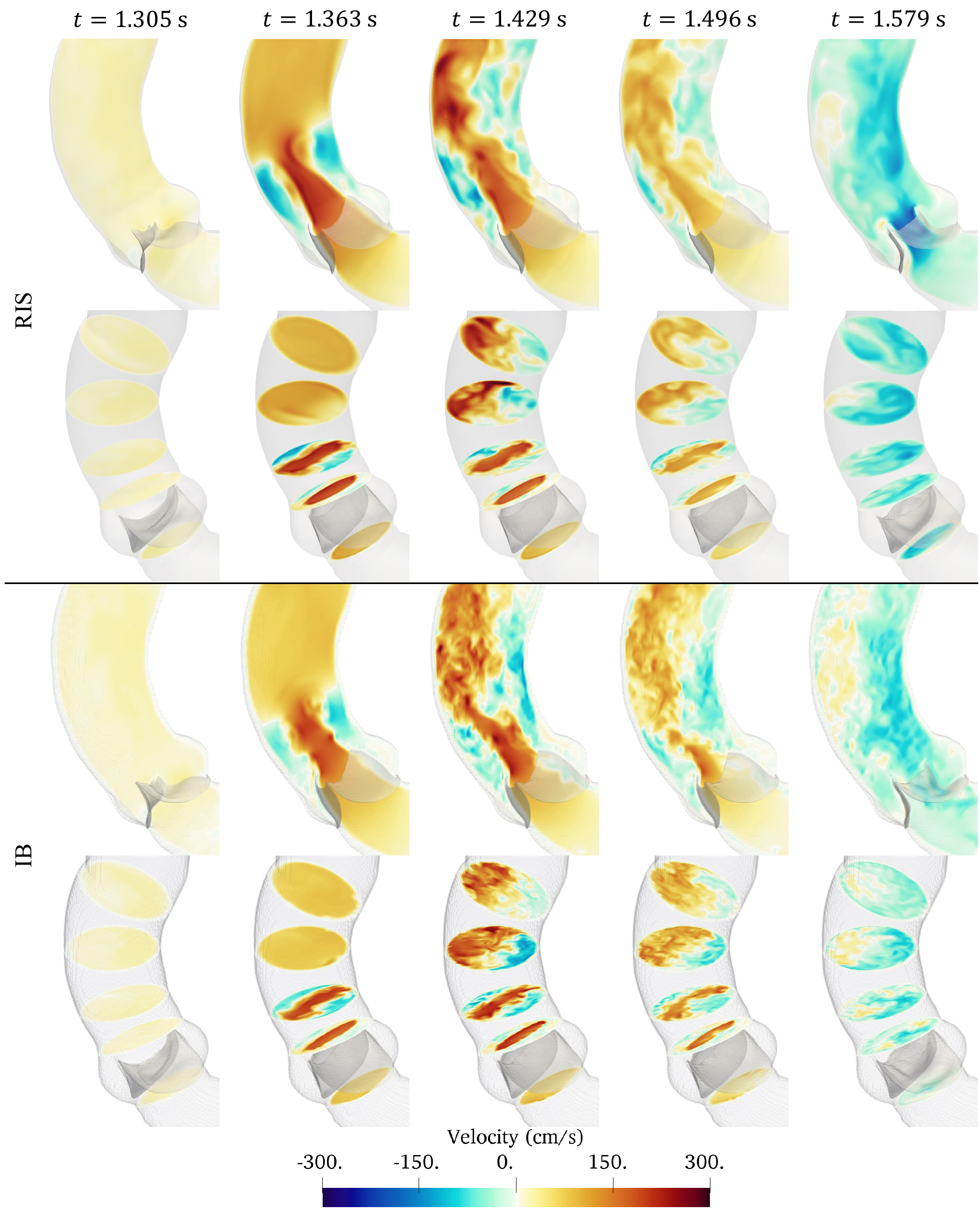}
    \caption{Comparison of the vertical velocity field and bulk flow structures in the bicuspid valve simulations obtained with the RIS and IB methods.}
    \label{fig:bicuspid-cfd}
\end{figure}

The simulation results for the bicuspid valve are shown in Figure~\ref{fig:bicuspid-cfd}. A trend similar to that of the trileaflet case is observed. The RIS and IB simulations produce very similar systolic bulk flow motion and comparable jet structures, while the IB results again retain more small-scale flow features. As discussed previously, this difference is associated with the specific discretization and temporal treatment used in the svMultiPhysics and IBAMR simulations, as well as the continued leaflet motion through fully coupled interaction after the valve has opened. Nevertheless, the overall flow pattern, including the systolic jet structure and the bulk flow evolution in multiple slices and at multiple time instants, is well captured by the RIS simulation.

For the first cardiac cycle, the RIS simulation gives a stroke volume close to the IB simulation, with 78.78~ml for RIS and 79.65~ml for IB. However, because of the larger backflow near the end of the first cardiac cycle in the RIS simulation, the chamber model produces a higher inlet pressure during the second cycle, which leads to a higher flow rate and stroke volume. In the second cardiac cycle, the stroke volumes are 89.10~ml for RIS and 76.22~ml for IB. Despite this difference, the mean pressure gradient predicted by RIS still agrees reasonably well with the IB result. In the first cardiac cycle, the mean pressure gradients are 8.64~mmHg and 8.42~mmHg for RIS and IB, respectively. In the second cycle, the mean pressure gradients are 10.38~mmHg for RIS and 12.75~mmHg for IB. The RIS method captures the overall magnitude of the pressure gradient. As in the trileaflet case, the RIS pressure gradient curves are more oscillatory and exhibit sharper changes, which are attributed to the rigid wall assumption and the prescribed leaflet motion.

\begin{figure}[!htb]
    \centering
    \hspace{\fill}
    \begin{subfigure}[t]{0.48\textwidth}
        \centering
        \includegraphics[width=\textwidth]{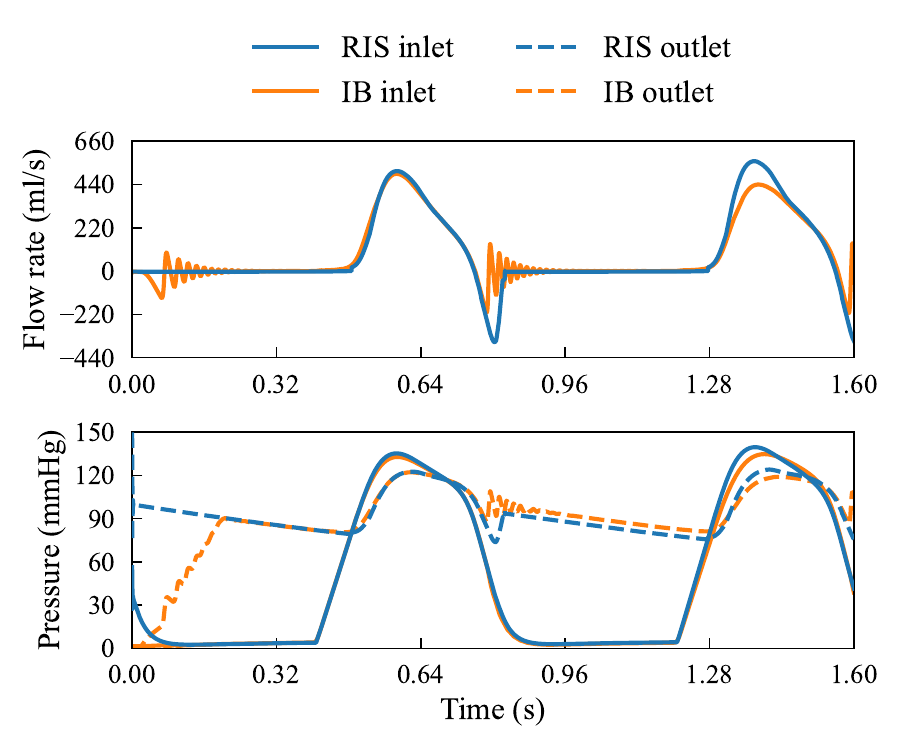}
        \caption{}
        \label{subfig:bicuspid-cfd-flowrate-pressure}
    \end{subfigure}
    \hspace{\fill}
    \begin{subfigure}[t]{0.48\textwidth}
        \centering
        \includegraphics[width=\textwidth]{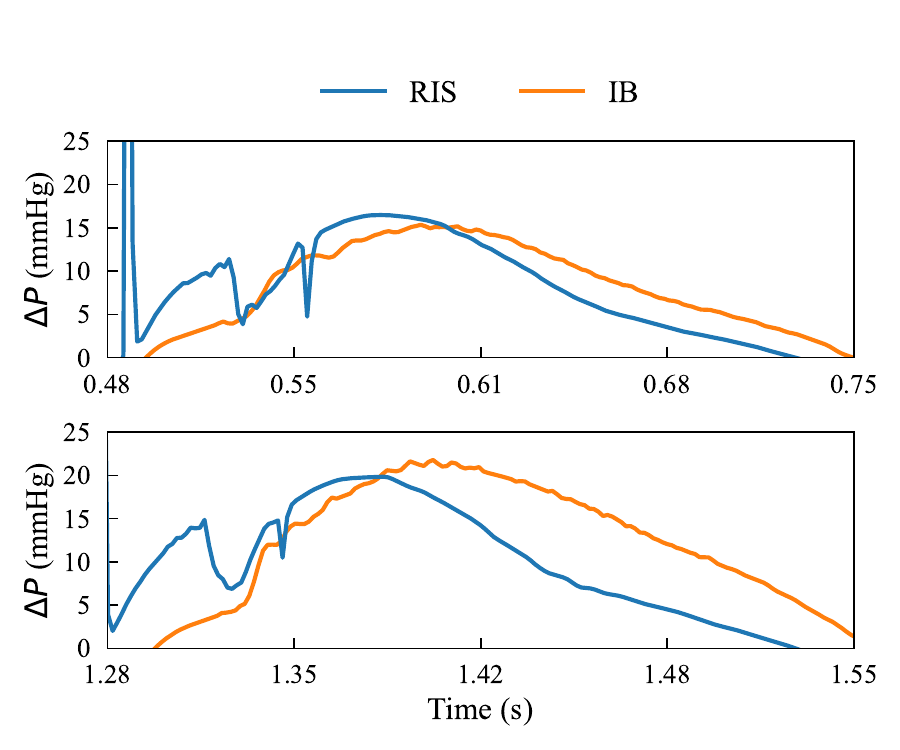}
        \caption{}
        \label{subfig:bicuspid-cfd-pressure-drop}
    \end{subfigure}
    \hspace{\fill}
    \caption{(a) Comparison of the flow rate and boundary pressure in the bicuspid valve simulations using the RIS and IB methods. (b) Comparison of the transvalvular pressure gradient $\Delta P$ predicted by the RIS and IB methods for the first and second cardiac cycles.}
    \label{fig:bicuspid-cfd-pressure-data}
\end{figure}

The bicuspid results show that the RIS framework can be applied to a different leaflet geometry and still recover the main systolic jet features and the overall order of magnitude of the transvalvular pressure gradient. However, the second-cycle discrepancies in stroke volume and pressure gradient reflect the same closure-driven feedback identified in the trileaflet case. The excess RIS backflow near the end of the first cycle alters the chamber-model inlet pressure for the subsequent cycle, which in turn shifts the second-cycle stroke volume. This indicates that the agreement between RIS and IB in cycle-to-cycle hemodynamics depends on how well the prescribed closing motion matches the IB closure dynamics. Therefore, this sensitivity is a general feature of the prescribed-kinematics RIS workflow.

\subsection{Aortic valve simulation with arterial wall using RIS method} \label{subsec:aortic-valve-fsi}

\begin{figure}[!htb]\centering
    \includegraphics[width=0.96\textwidth]{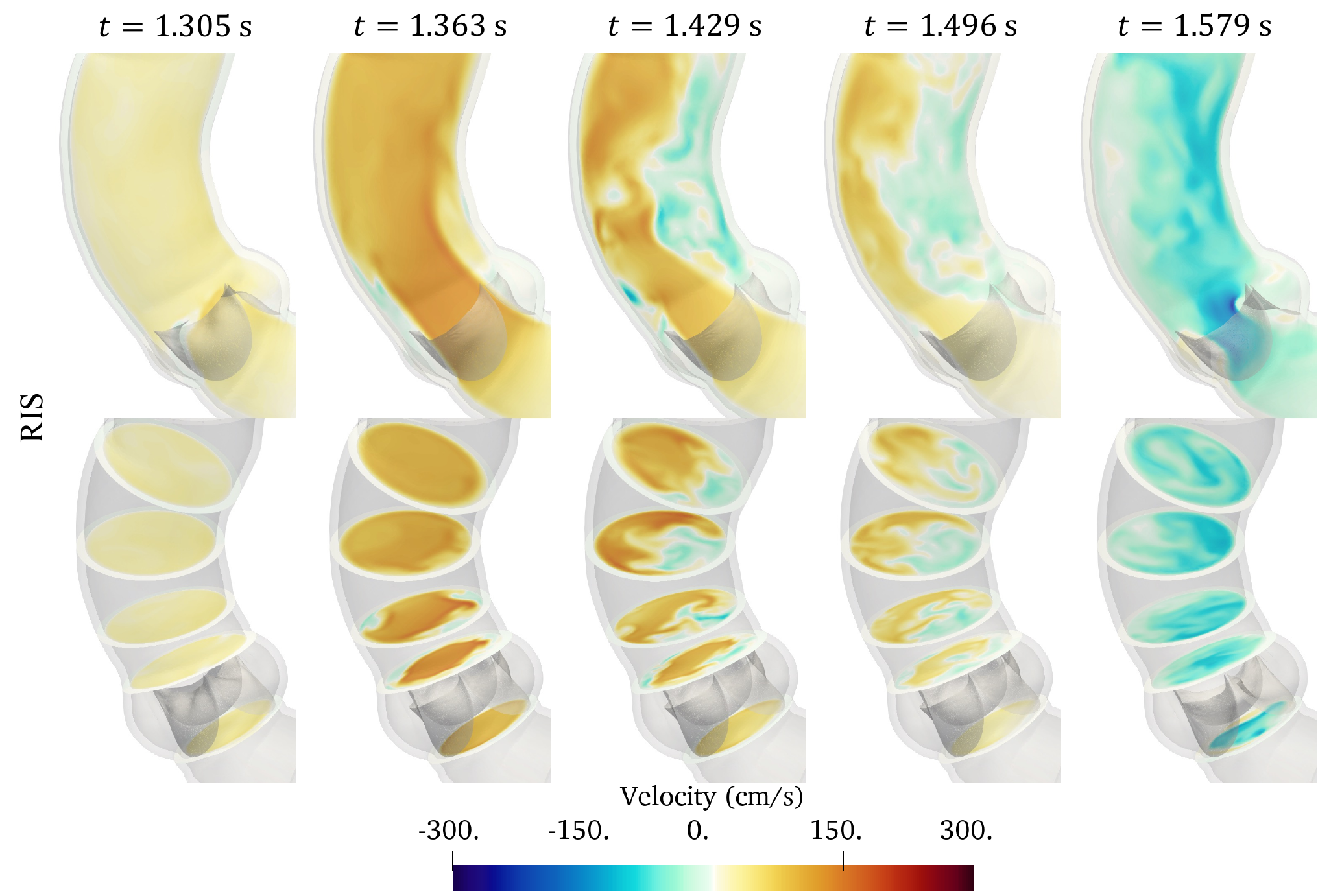}
    \caption{Vertical velocity field in the aorta obtained from the RIS simulation with an elastic arterial wall.}
    \label{fig:trileaflet-fsi}
\end{figure}

In this section, the aortic valve simulation is performed using the RIS method and the prescribed leaflet motion shown in Figure~\ref{fig:ris-trileaflet-motion}, but with an elastic arterial wall included in the simulation. This additional case is presented to illustrate an extension implemented in svMultiPhysics rather than a strictly equivalent comparison with the rigid wall IB baseline. The arterial wall is modeled with a thickness of 2~mm and a mesh size of 1~mm, using four layers of elements across the wall thickness. The inner surface mesh of the arterial wall conforms to the outer surface of the aortic fluid mesh used in the RIS simulation. The FSI between the aortic flow and the arterial wall is therefore solved in a monolithic manner. The wall motion is handled within the ALE framework, where the mesh deformation is governed by an elasticity problem. The arterial wall is modeled using a Neo-Hookean material model and is constrained by Dirichlet boundary conditions at the inlet and outlet, together with Robin boundary conditions on the outer wall surface. The material parameters of the arterial wall are a Poisson’s ratio of 0.45, an elastic modulus of $1.0 \times 10^7~\mathrm{dyne}/\mathrm{cm}^2$, and a density of $1.0~\mathrm{g}/\mathrm{cm}^3$. The stiffness and damping coefficients in the Robin boundary condition are $1.0 \times 10^6~\mathrm{dyne}/\mathrm{cm}^3$ and $1.0 \times 10^3~\mathrm{dyne}\cdot \mathrm{s}/\mathrm{cm}^3$, respectively. The arterial wall contributes an additional 453,501 DoFs to the coupled system. All other simulation settings and boundary conditions remain unchanged. Representative fluid velocity fields are shown in Figure~\ref{fig:trileaflet-fsi}.

\begin{figure}[!htb]
    \centering
    \hspace{\fill}
    \begin{subfigure}[t]{0.48\textwidth}
        \centering
        \includegraphics[width=\textwidth]{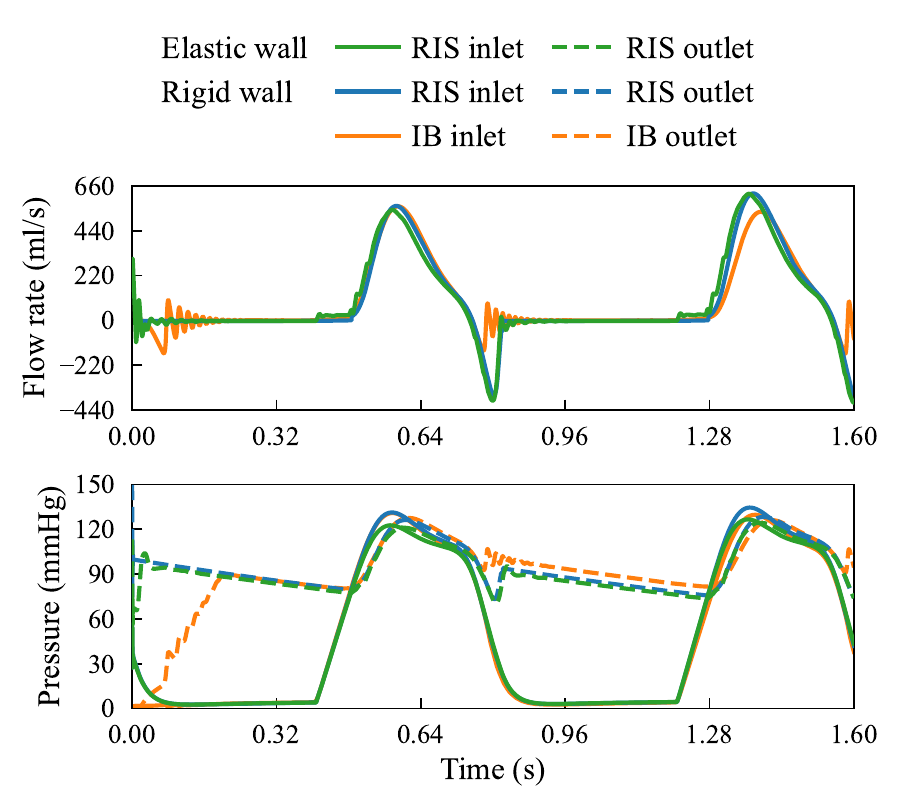}
        \caption{}
        \label{subfig:trileaflet-fsi-flowrate-pressure}
    \end{subfigure}
    \hspace{\fill}
    \begin{subfigure}[t]{0.48\textwidth}
        \centering
        \includegraphics[width=\textwidth]{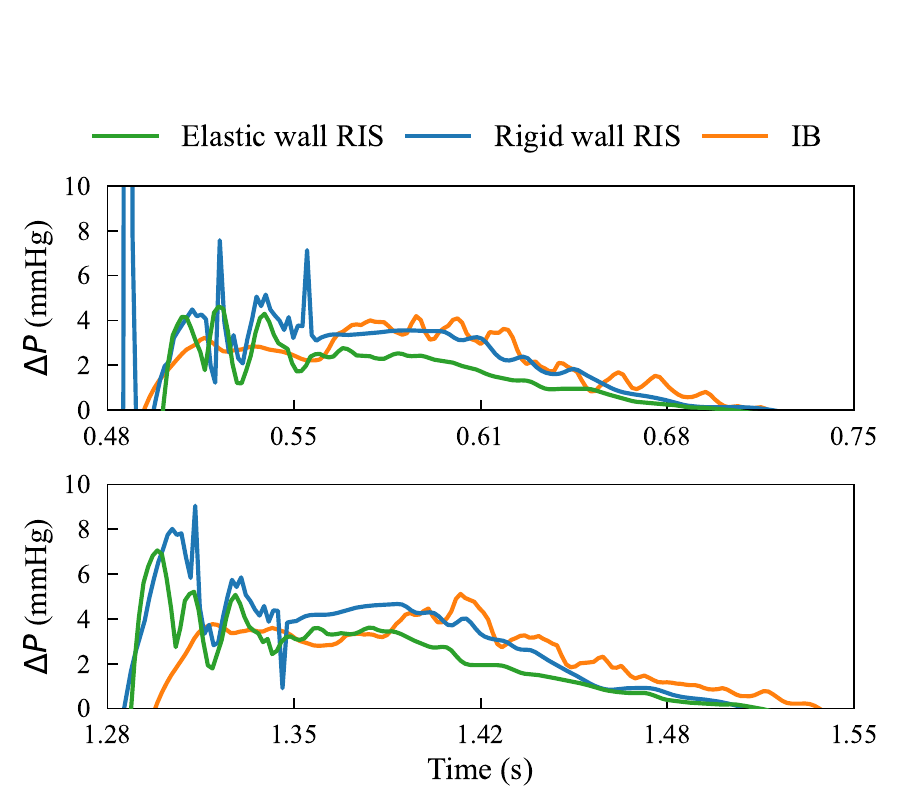}
        \caption{}
        \label{subfig:trileaflet-fsi-pressure-drop}
    \end{subfigure}
    \hspace{\fill}
    \caption{(a) Comparison of the flow rate and boundary pressure between the elastic wall RIS, rigid wall RIS, and IB simulation for trileaflet aortic valve. (b) Comparison of the transvalvular pressure gradient $\Delta P$ predicted by the elastic wall RIS, rigid wall RIS, and IB simulations for the first and second cardiac cycles.}
    \label{fig:trileaflet-fsi-pressure-data}
\end{figure}

The overall flow pattern for the RIS aortic valve simulation with an elastic arterial wall remains similar to that of the rigid wall RIS simulation. However, the flow speed is slightly lower in the moving wall case. During systolic expansion, the compliant wall increases the local cross-sectional volume, so part of the incoming fluid volume is accommodated by the enlarged vessel, reducing the instantaneous flow speed. Wall motion also affects the flow separation and jet development after valve opening, as the outward expansion of the wall widens the lumen and alters the trajectory and extent of the jet, which can be observed in the third snapshot of Figure~\ref{fig:trileaflet-fsi}. Even with these differences, the flow rate curve and inlet pressure profile remain close to those of the IB simulation, with trends similar to those observed in the rigid wall RIS case, as shown in Figure~\ref{subfig:trileaflet-fsi-flowrate-pressure}. A slight oscillation is observed in the elastic wall flow rate curve at valve opening due to the interaction between the prescribed leaflet motion and the elastic wall, and it damps out quickly without affecting the overall flow behavior.

Furthermore, the mean transvalvular pressure gradient is 1.76~mmHg and 2.32~mmHg for the first and second cardiac cycles, respectively. The slightly lower mean pressure gradients, compared with 2.23~mmHg and 2.47~mmHg from the rigid-wall IB simulation, reflect a local effect of wall compliance near the valve. With a rigid wall, the abrupt fluid accelerations and decelerations induced by the prescribed RIS leaflet motion during opening and closure generate sharp pressure transients, since the local volume cannot adjust to accommodate them. These transients appear directly in the pressure field. With a compliant wall, the aortic segment near the valve expands and contracts slightly in response, absorbing part of these pressure oscillations. This effect lowers both the peak and average transvalvular pressure gradient and is distinct from the effect produced by the outlet RCR model. In addition, the pressure gradient curves are significantly smoother than those obtained in the rigid wall RIS simulation. This improvement is due to arterial wall compliance, which reduces abrupt pressure variations and mitigates the oscillatory response associated with the rigid wall configuration. In the rigid wall RIS simulation, the flow must adjust more abruptly to the prescribed leaflet motion, thereby enhancing oscillations in the pressure field.

The inclusion of an elastic arterial wall provides an additional benefit in the RIS simulation by reducing oscillatory pressure behavior and avoiding abrupt pressure changes. This is particularly beneficial in the present RIS setting, where the leaflet motion is prescribed and remains relatively rigid during opening and closure.

\section{Discussion} \label{sec:discussion}

In this study, we compared a prescribed kinematics RIS workflow in svMultiPhysics with a fully coupled IB workflow in IBAMR for trileaflet and bicuspid aortic valve configurations. Because the RIS leaflet motion was extracted from the IB simulations, the comparison is intended to evaluate how well RIS can reproduce hemodynamic quantities of interest when kinematic input is available, rather than to establish equivalence between RIS and IB as fully independent predictive valve simulators. In this controlled comparison, the RIS results captured the main systolic flow patterns and the overall magnitude of the mean transvalvular pressure gradient, while discrepancies remained in closure-sensitive quantities and in cycle-to-cycle responses induced by the lumped-parameter boundary conditions. The fixed-valve test suggests that part of the flow discrepancy is associated with solver and discretization choices, whereas the moving-valve cases reflect the difference between prescribed RIS motion and fully coupled IB leaflet dynamics.

The RIS method offers several practical advantages. Because the valve kinematics are prescribed, it requires only the solution of the fluid equations and does not involve a leaflet structural solve or iterative fluid–structure coupling, which reduces both modeling and implementation complexity. In svMultiPhysics, adding the RIS valve introduces only a modest increase in computational cost. To quantify this overhead, we ran a baseline aorta CFD simulation with the same mesh and boundary conditions but without the valve. The baseline simulation takes 6 hours 24 minutes per cardiac cycle on 48 processors, while the same simulation with the RIS valve takes 7 hours 55 minutes under the same boundary conditions. In contrast, the fully coupled IB simulation requires 20 hours 51 minutes per cardiac cycle on the same number of processors. All simulations were performed on Stanford’s Sherlock HPC cluster across 2 Intel Xeon Gold 5118 CPU nodes (24 cores per node). This indicates that the RIS valve does not significantly increase the overall cost of the simulation compared to the baseline aorta CFD simulation, while the fully coupled IB simulation requires approximately 2.6 times more wall-clock time. The second advantage is that the RIS formulation integrates naturally with the ALE FSI framework in svMultiPhysics, allowing coupling with elastic arterial wall models without changing the valve treatment. Lastly, the kinematic input to RIS is not restricted to IB-derived data; it can also be obtained from medical imaging~\cite{fedele2017patient} or from structural analyses, making RIS attractive for clinical or large-scale computational pipelines in which a full FSI simulation is unavailable or impractical.

For the aortic valve simulation with an elastic arterial wall discussed in Section~\ref{subsec:aortic-valve-fsi}, the RIS method coupled with the ALE FSI framework requires an explicit geometric coupling strategy~\cite{bucelli2022partitioned} to achieve convergence. In this strategy, the monolithic fluid--structure system is first solved to convergence, and the mesh equation is then solved separately before advancing to the next time step. The standard implicit coupling strategy, where the fluid--structure and mesh equations were iterated alternately until convergence, showed convergence difficulties when the RIS force term is included.

The primary limitation of the RIS approach is that the valve kinematics must be provided as input, so leaflet motion is prescribed rather than computed. Because the closing motion is prescribed rather than driven by the fluid loading, the RIS closure dynamics are not consistently coupled to the instantaneous flow state, which results in a larger reverse flow spike at closure than in the IB simulation, although this spike can be reduced by adjusting the prescribed closing speed and timing. The excess backflow propagates into the next cardiac cycle through the lumped-parameter model, producing a larger second cycle stroke volume and systolic flow rate than in the IB simulation. The RIS method is therefore not suitable for quantitative prediction of regurgitant volume or sealing performance. In addition, due to the absence of the structural solver for the valve leaflets, the RIS method provides no information on leaflet stresses, strains, or structural loading. Studies targeting valve mechanical performance, fatigue life, or structural integrity require a method that resolves the two-way coupled FSI, such as the IB method.

The prescribed inlet pressure test in Section~\ref{subsec:aortic-valve-comparison} provides additional insight into which discrepancies are attributable to the RIS valve formulation and which are caused by feedback effects from the lumped-parameter boundary conditions. When the inlet pressure waveform extracted from the IB simulation is imposed directly, the forward flow rates and mean transvalvular pressure gradients agree more closely between the two methods across both cardiac cycles. This indicates that the large second cycle stroke volume difference observed in the original chamber model comparison is primarily a boundary-condition feedback effect. In RIS, the excess backflow drives the inlet pressure higher in the subsequent cycle. To assess whether this divergence continues beyond the second cycle, an additional RIS simulation is extended to three cardiac cycles. The third cycle flow rate and pressure responses are nearly identical to those of the second cycle, indicating that the RIS simulation reaches a cycle-to-cycle periodic state by the second cycle rather than continuing to diverge.

A practical consideration in applying RIS is the sensitivity to the two user-specified parameters: the resistance coefficient $R$ and the half-thickness $\varepsilon$. In the present simulations, $R=1.0\times10^5~\mathrm{g}/(\mathrm{cm}\cdot \mathrm{s})$ provides strong suppression of through surface flow without causing obvious conditioning issues. Following prior RIS studies~\cite{fedele2017patient}, $\varepsilon$ is selected as approximately 1.5 times the local mesh size. These two parameters have distinct effects on the valve behavior. For the resistance coefficient, a value of $R$ that is too small allows fluid to pass through the leaflet surface, resulting in unphysical leakage and an increased flow rate. In contrast, an overly large $R$ makes the discrete system ill-conditioned. For the half-thickness, a larger $\varepsilon$ reduces the effective orifice area, thereby increasing the transvalvular pressure gradient and decreasing the flow rate. However, $\varepsilon$ cannot be made arbitrarily small. It should span approximately 1.5 surrounding fluid element widths so that the smoothed Dirac delta function in Eq.~\eqref{eq:ris_dirac_delta_function} produces a well-resolved resistive force distribution.

The workflow that uses IB kinematics as input to RIS is particularly useful in scenarios where a single high-fidelity IB simulation is first used to generate representative kinematics, after which RIS can be used for downstream studies such as parameter sweeps, cohort studies requiring repeated simulations with varied geometries, or rapid hemodynamic screening. In addition, RIS kinematics are not restricted to IB simulations. In clinical and engineering settings, they can be derived from medical imaging data or structural analyses, both of which avoid fully coupled FSI. In these cases, RIS provides an efficient approach for estimating bulk hemodynamic quantities, such as flow structures and transvalvular pressure gradients, without the cost of two-way coupling. The present IB and RIS comparison therefore provides a controlled assessment of RIS accuracy. Overall, the choice between the two methods should be guided by the target quantities of interest and the availability of kinematic input data.

\section{Conclusion} \label{sec:conclusion}
This work compared a prescribed-kinematics RIS formulation implemented in svMultiPhysics with a fully coupled IB formulation implemented in IBAMR for trileaflet and bicuspid aortic valve simulations. When leaflet kinematics were provided from the IB reference, the RIS approach reproduced the main large-scale systolic flow features and transvalvular pressure gradient. Combining the RIS model with an elastic arterial wall reduced the oscillatory character of the pressure gradient response relative to the rigid wall RIS simulation. The largest discrepancies occurred during valve closure and in the subsequent coupled cycle response, where the prescribed RIS motion could not adapt to instantaneous fluid loading. In summary, RIS provides a reliable representation of bulk hemodynamics, including flow structures, flow rates, and pressure gradients, under matched boundary conditions. Compared to the fully coupled IB simulation, it achieves significantly reduced computational cost while maintaining comparable accuracy in the resolved flow features. This makes RIS a computationally efficient alternative for applications where capturing hemodynamic influence of the valve is the primary goal and fully two-way coupled FSI is not required.

\section*{Acknowledgments}
HZ, DC, and SD were supported by the National Science Foundation under award number 2310909. HZ and ALB were supported in part by National Institutes of Health (NIH), United States of America grants 5R01HL129727-08 and 5R01HL159970-04. ADK was supported in part by a grant from the National Heart Lung and Blood Institute of the National Institute of Health under award number K25HL175208, and American Heart Association grant 24CDA1272816 (https://doi.org/10.58275/AHA.24CDA1272816.pc.gr.193564) and a grant from Stanford Maternal and Child Health Research Institute. FK was supported by the National Science Foundation award number 1663671, and the National Institute of Health award number R01EB029362. ZH was supported by Additional Ventures Cures Collaborative. 
We thank Dr. Michele Bucelli at the University of Texas at Austin for discussions on explicit geometric coupling. Computing for this project was performed on the Sherlock cluster at Stanford University with assistance from the Stanford Research Computing Center. 

\bibliography{main}

\end{document}